\documentclass[a4paper,fleqn,usenatbib]{mnras}

\usepackage[T1]{fontenc}
\usepackage{ae,aecompl}

\usepackage{multirow}
\usepackage{mathtools}

\usepackage{graphicx}	
\usepackage{amsmath}	
\usepackage{amssymb}	

\usepackage{wasysym}

\title[The GALAH survey: The data reduction pipeline]{The GALAH survey: The data reduction pipeline}

\author[J. Kos et al.]{
Janez Kos,$^{1,2}$\thanks{E-mail: jkos@physics.usyd.edu.au}
Jane Lin,$^{3}$
Toma\v{z} Zwitter,$^{2}$
Maru\v{s}ka \v{Z}erjal,$^{2}$
Sanjib Sharma,$^{1}$\newauthor
Joss Bland-Hawthorn,$^{1}$
Martin Asplund,$^{3}$
Andrew R. Casey,$^{4}$\newauthor
Gayandhi M. De Silva,$^{5,1}$
Ken C. Freeman,$^{3}$
Sarah L. Martell,$^{6}$\newauthor
Jeffrey D. Simpson,$^{5}$
Katharine J. Schlesinger,$^{3}$
Daniel Zucker,$^{7,8}$\newauthor
Borja Anguiano,$^{7}$
Carlos Bacigalupo,$^{7}$
Timothy R. Bedding,$^{1}$\newauthor
Christopher Betters,$^{1,9}$
Gary Da Costa,$^{3}$
Ly Duong,$^{3}$
Elaina Hyde,$^{10}$\newauthor
Michael Ireland,$^{3}$
Prajwal R. Kafle,$^{11}$
Sergio Leon-Saval,$^{1,9}$
Geraint F. Lewis,$^{1}$\newauthor
Ulisse Munari,$^{12}$
David Nataf,$^{3}$
Dennis Stello,$^{1}$
Chris G. Tinney,$^{5,13}$\newauthor
Gregor Traven,$^{2}$
Fred Watson,$^{5}$
Robert A. Wittenmyer$^{5,13}$\\
\\
$^{1}$Sydney Institute for Astronomy, School of Physics, A28, The University of
Sydney, NSW, 2006, Australia\\
$^{2}$Faculty of Mathematics and Physics, University of Ljubljana, Jadranska 19, 1000 Ljubljana, Slovenia\\
$^{3}$Research School of Astronomy \& Astrophysics, Australian National University, ACT 2611, Australia\\
$^{4}$Institute of Astronomy, University of Cambridge, Madingley Road, Cambridge CB3 0HA, UK,\\
$^{5}$Australian Astronomical Observatory, North Ryde, NSW 2133, Australia\\
$^{6}$School of Physics, UNSW, Sydney, NSW 2052, Australia\\
$^{7}$Department of Physics and Astronomy, Macquarie Univeristy, Sydney, NSW 2109, Australia\\
$^{8}$Research Centre in Astronomy, Astrophysics \& Astrophotonics, Macquarie University, Sydney, NSW 2109, Australia\\
$^{9}$Sydney Astrophotonic Instrumentation Labs, School of
Physics, A28, The University of Sydney, NSW 2006, Australia\\
$^{10}$Western Sydney University, Locked Bag 1797, Penrith, NSW 2751, Australia\\
$^{11}$International Centre for Radio Astronomy Research (ICRAR), The University of Western Australia, 35 Stirling Highway, \\Crawley, WA 6009, Australia\\
$^{12}$INAF National Institute of Astrophysics, Astronomical Observatory of Padova, 36012 Asiago, Itay\\
$^{13}$Australian Centre for Astrobiology, UNSW Australia, Sydney, NSW 2052, Australia\\
}

\date{Accepted XXX. Received YYY; in original form ZZZ}

\pubyear{2016}

\begin{document}
\label{firstpage}
\pagerange{\pageref{firstpage}--\pageref{lastpage}}
\maketitle

\begin{abstract}
We present the data reduction procedures being used by the GALAH survey, carried out with the HERMES fibre-fed, multi-object spectrograph on the 3.9~m Anglo-Australian Telescope. GALAH is a unique survey, targeting 1 million stars brighter than magnitude V=14 at a resolution of 28,000 with a goal to measure the abundances of 29 elements. Such a large number of high resolution spectra necessitates the development of a reduction pipeline  optimized for speed, accuracy, and consistency. We outline the design and structure of the Iraf-based reduction pipeline that we developed, specifically for GALAH, to produce fully calibrated spectra aimed for subsequent stellar atmospheric parameter estimation. The pipeline takes advantage of existing Iraf routines and other readily available software so as to be simple to maintain, testable and reliable. A radial velocity and stellar atmospheric parameter estimator code is also presented, which is used for further data analysis and yields a useful verification of the reduction quality. We have used this estimator to quantify the data quality of GALAH for fibre cross-talk level ($\lesssim0.5$\%) and scattered light ($\sim5$ counts in a typical 20 minutes exposure), resolution across the field, sky spectrum properties, wavelength solution reliability (better than $1$~$\mathrm{km\ s^{-1}}$ accuracy) and radial velocity precision.
\end{abstract}




\section{Introduction}

There is no definitive recipe for reducing multi-fibre spectra.
Multi-fibre, multi-objects spectrographs are sophisticated and complex
instruments. The complexity of these instruments precludes the analysis of its data by individual users on their own, so the users rely on reduction pipelines provided by the
instrument team. To date each such instrument \citep[e.g.][]{blecha00,palmero14,luo15,nidever15} has developed a dedicated data-processing pipeline that delivers processed spectra for that instrument to its users (including large surveys being carried out with the instrument). The quality of
any scientific analysis depends strongly on the quality of the
reduced spectrum. It is critical that users understand the details of the reduction pipeline, such as the reduction process, steps made, and respective levels of calibration and the accuracy of the delivered data to take a full advantage of the products.

The GALAH (GALactic Archeology with Hermes) survey \citep{desilva15} uses the HERMES fibre-fed, multi-object spectrograph at 3.9~m Anglo-Australian Telescope (AAT) -- itself fed by the 2dF fibre positioner system. The spectrograph delivers high resolution spectra in four wavelength ranges. By combining multi-object capability with a high resolution spectrograph, the GALAH survey aims to measure the
abundances of 29 elements for one million stars over a five year period. GALAH is the first major survey being conducted with this spectrograph which was designed with survey goals in mind \citep{sheinis15}. The fibre positioner system (2dF), however, is older and used by several other multi-fibre spectrographs -- primarily for surveys targeting extragalactic sources \citep[e.g.][]{colless03, croom04, drinkwater10}. These have all made use of a common reduction pipeline called 2dfdr \citep{aao15}. 2dfdr serves several different spectrographs that all use the common 2dF fibre positioner system. A reduction pipeline within the 2dfdr environment was also
written for HERMES, but it was not suitably optimised for GALAH (unreliable wavelength calibration) and lacked many important features (sophisticated sky removal algorithm), so it often produced results of poor quality. Implementation of features 2dfdr lacks is easy in Iraf, as it is more general purpose software and gives the user options that have not been considered yet in 2dfdr.

We have therefore developed an alternative pipeline for the GALAH survey. It is based on pre-existing Iraf\footnote{IRAF is distributed by the National Optical Astronomy Observatory, which is operated by the Association of Universities for Research in Astronomy (AURA) under a cooperative agreement with the National Science Foundation.} routines in order to be simple to maintain. Iraf is convenient to use, because more time and care can be invested into building functional and advanced routines that are constitute a pipeline instead of writing the functions themselves that already exist in the Iraf distribution. Iraf functions, however, are not enough to create a pipeline. They serve a very general pourpose and we use it as any library of computer code in the process described in this paper. Iraf has been widely used in the astronomy community for more than two decades. The routines we use have therefore been extensively tested and proved to be reliable \citep{tody86}. Our Iraf pipeline is now the default pipeline used in the GALAH survey. While it is regularly updated and improved, it remains an internal project, and is not intended for  general HERMES users. Some results and analysis of the HERMES spectrograph, on the other hand, serve a much broader audience and are not specific to the GALAH survey. The same holds for some routines we develped (like cosmic removal from images with a strong patern, fitting and removing sky spectra, and callibration of the Thorium-Xenon arc lamp) that serve all HERMES users and other spectrographs that are dealing with the same problems.

This paper briefly reviews observational aspects of the GALAH survey in Section \ref{sec:procedure} and the motivation for a new reduction pipeline in Section \ref{sec:mot}. The main part of the paper is a step-by-step description of the pipeline (Section \ref{sec:pipeline}). In Section \ref{sec:org}, we describe the organization of reduced data. As a part of the reduction pipeline we also run another code, a simple estimator of the radial velocity and basic parameters of stellar atmospheres. This is discussed in Section \ref{sec:rv}. Two approaches to measuring the resolution of the GALAH spectra are presented in Section \ref{sec:ress}. An important by-product of this work -- the line lists for the wavelength calibration lamp are collected in the appendix.

\section{Observing procedure}
\label{sec:procedure}
The observing procedures for the GALAH survey are described in detail elsewhere \citep{sarah2016,sanjib2016}, so we only summarise relevant parts here. 

HERMES is a high-resolution, multi-fibre spectrograph, simultaneously observing four wavelength ranges in different arms (see Table \ref{tab:wav}). The resolving power is 28,000 in normal observing mode (used for the GALAH survey) or 45,000 in a slit-mask mode. The resolution in the GALAH observing mode corresponds to a resolution element size of ~3.5 pixels on each of the four CCDs. Note that due to the design of the spectrograph, the wavelength ranges vary by a few {\AA}ngstroms with the position of the spectrum on the CCD.

\begin{table}
\centering
\begin{tabular}{lcc}
\hline\hline
Arm & $\lambda_{\mathrm{min}}$ / \AA & $\lambda_{\mathrm{max}}$ / \AA\\\hline
Blue & 4718 & 4903\\
Green & 5649 & 5873\\
Red & 6481 & 6739\\
IR & 7590 & 7890\\\hline
\end{tabular}
\caption{Wavelength ranges covered by each arm of the HERMES spectrograph.}
\label{tab:wav}
\end{table}

The GALAH survey fields are defined by a fixed pattern across the southern sky \citep{sanjib2016}. Each field has a unique observing configuration file with information for each star within its sky area. The observer selects a field from a list of observable fields at a given time and feeds it into the target allocation software. This software calculates the optimal fibre arrangement to observe as many targets in the field as possible. It also allocates sky fibres and selects guide stars. Once the observer is satisfied with the configuration, it is sent to the 2dF fibre positioner robot, which re-arranges the fibres into the correct positions.

Fibres are held in position with magnets on one of two metal plates (called plate 0 and plate 1). These face away from each other, such that the fibres can be positioned on one plate, while a separate set of fibres are receiving light from the telescope on the other plate. 

There are 400 fibres on each plate available to the 2dF positioner robot. Of these, 392  are regular fibres and 8 are fiducial (auxiliary) fibres used for guiding. On the telescope end, fibres on each plate are given a so called pivot number from 1 to 400. Bundles of 10 fibres are then fed into the spectrograph where they form a pseudo-slit. Each such group of 10 fibres is called a slitlet. In each slitlet is an optical element that changes the focal ratio of the fibres. This inverts the numbering order within each slitlet. The fibres on the spectrograph end are again numbered from 1 to 400 (now called fibre numbers) but, because of the inversions, fibre numbers do not correspond directly to pivot numbers. During the reduction we ignore the fiducial fibres (they do not contribute any signal to the image), so we only have to deal with 392 spectral traces. When apertures are fitted to the traces they are again numbered (this time from 1 to 392), so these aperture numbers do not correspond to either fibre or pivot numbers. Conversion between pivot and fibre numbers is given in the table A1 in \citet{jeffrey2016}. A further conversion between fibre and aperture numbers is simple: the aperture number follows the same order as the fibre number, but skips 8 fiducial fibres. Fire number one is therefore aperture number one, but fibre number 400 is aperture number 392.

At the beginning of each night, observers take a series of bias frames for each CCD. These are used for the reduction of the science, flat and arc lamp images and do not have to be repeated later in the night. At the same time, the first two fields are sent to the fibre configuration software and the fibres are positioned on both plates. After the first field is observed, the plates are switched and the second field is observed, while another field is configured on the now available first plate. 

Each field is usually exposed three times for 20 minutes per exposure. The observer can decide if more exposures are needed or if they should be aborted due to bad weather, for example. One exposure of the arc lamp and one exposure of the flat lamp are made with the same fibre configuration as for the science exposures. Not all fields require 1 hour of exposure time. During twilight we observe bright fields that require less exposure time, while special programs (such as observations of standard stars or K2 follow-up) also use different exposure times.

The images created by the four CCDs (one per spectrograph arm) are kept in separate folders and are identified by their run number, which follows the consecutive exposure done in that night. The fibre configuration used (including sky, fiducial, and unused fibres) is also saved -- to record which star was positioned in which fibre.

The rules, exceptions and patterns described above guided the design of the reduction pipeline.

\section{Motivation}
\label{sec:mot}
It was originally intended that the GALAH survey data should be processed by the AAO's 2dfdr software. However, the software is designed for many different instruments with a wide range of data types. Thus, 2dfdr lacked specific functionality required to cope with problems unique to high-resolution spectroscopy. Thus a more efficient and more specialized code was needed. A decision was made to write a second pipeline, produced and maintained by the GALAH collaboration for the purpose of the survey. Such a pipeline is under the control of the GALAH team, so any adjustments can be implemented quickly. The pipeline is fine-tuned to work only with GALAH spectra that have been taken in a consistent predictable format. It is not meant to be released as a pipeline for general use. 

To ensure the pipeline was fast to develop, easy to maintain or modify and reliable, we have made extensive use of pre-existing and already tested Iraf routines. For procedures not available in the Iraf packages, we, again, used existing solutions before writing our own. Due to its foundation in Iraf, we refer to this pipeline simply as the ''Iraf pipeline''. Because the pipeline uses readily available software, the whole pipeline contains only $\sim$2000 lines of code. The cost of such approach is a slow execution of the code. The reduction (including radial velocity and stellar atmospheric parameter estimation) of the current GALAH dataset (at the time of writing 200,000 unique stars or 600 fields or around 100 nights assuming the maximum efficiency) takes almost 2 weeks on a top-end desktop PC. The main limitation is that Iraf uses many hard drive read/write operations on files at numerous steps. In spite of the speed, Iraf had many advantages that suited our requirements.

We have striven to make the pipeline automatic, requiring no human interaction to reduce the data. However, for some steps, complete automation requires disproportionate effort to implement, so human intervention is needed when observing strategy is changed, or spectrograph properties change.

It was also felt to be desirable that the same code is used to reduce spectra from all four arms to simplify the pipeline significantly, which has been achieved in the pipeline presented here, except for a few steps that are skipped entirely for spectra from some arms because they are unnecessary.

\section{The pipeline}
\label{sec:pipeline}
The major stages used by the pipeline are presented here in a logical order, but note that they are not
necessarily executed in the same order, due to complications
like an individual step requiring multiple iterations.

The Iraf pipeline reduces data on a night-by-night basis, and
the same procedure is used for all images, regardless of which
observing program produced them. Before submission to the pipeline, images are separated into 8 groups -- one for each of the four arms in either of the two plates. Each group of images is then reduced on its own without any interaction with the processing for other groups. Images from each night are reduced separately from all other nights. Any number of executions of the pipeline can be run in parallel, making the reduction process faster and more efficient. Usually we run between 16 and 32 jobs at once, so 2 to 4 nights can be reduced in parallel. 

\subsection{Non-automated routines}
\label{sec:nonauto}
As noted earlier, some pipeline steps cannot be completely automated. They do not have to be performed every time the pipeline is run -- some are done once for all nights, while others are updated from time to time -- usually as a result of a changed instrument calibration or a new observing strategy.

Spectrum traces (see Figure \ref{fig:strike} for an example of how raw HERMES spectra look like) are found by taking a cross-section along a specified CCD column and locating the peaks representing the traces. The trace-finding algorithm relies heavily on a ''key'' which indicates which aperture is positioned at which pixel. The position and separation between traces are not uniform and can also change when cameras are refocused or CCDs moved. A key was produced that provides the location of the trace centre for each aperture in a reference column. When traces are found in a new image, a global shift is allowed against the key and a small shift of 3 pixel for each individual aperture is also allowed. In comparison, the full width at half maximum (FWHM) of the trace is around 4 pixels in the spatial direction and the aperture width used for spectral extraction is 6 pixels. The key has to be generated by hand and can be used for as long as the geometry of the traces does not change significantly. At the time of writing this has only been done once. 

Once the traces are identified in the reference column, they are propagated to other columns with the same criterion that the deviation from the key is of 3 pixels at most. This guarantees that the apertures are positioned on the correct spectra every time and that the trace does not jump between the neighbouring spectra.

Wavelength calibration also requires some human intervention. A lot of time has been invested in building the line lists for the wavelength calibrations (see Section \ref{sec:wav} for a detailed description). The line lists alone, however, are not enough to make a wavelength calibration. The method also expects to be given an approximate relationship to transform pixels into wavelengths with sufficient precision such that correct  arc lines are identified. The wavelength calibration was initially done manually for a large number of spectra, which allowed this relationship to be established. Humans are much more likely to notice mismatched lines than computers and the established relationship is used as a proxy for the wavelength solution. We monitor the final wavelength solution from time to time to determine if the proxy relationship has to be updated for new data.

We use human verification to assist with cosmic ray removal. An important parameter that was found by eye is the threshold in the cosmic ray removal algorithm. If this removal threshold is set too low, useful data will be tagged as cosmic rays, significantly impacting the quality of the spectra. If set too high, the cosmic rays will not be removed. Different types of images (normal exposures, fibre configurations with only one bright star, exposures with low signal due to cloud cover, etc.) were run through the cosmic ray removal algorithm and inspected by eye. The thresholds chosen as a result were then hard-coded to give the optimal result in every type of image we checked. 

Finally, a large number of parameters (e.g. orders of fitted polynomials) were also determined by reducing a few images by hand and adjusting the parameters to get the best results. They are explained in their appropriate sections below.

\subsection{Cosmetic corrections}
With each set of science frames, we take one flat frame and one arc frame, either before or after the science frames. The flat frame is a so-called flap flat, produced by a system of lights shining on flaps inserted into the telescope below the prime focus location of the 2dF, that direct lamp light onto the fibre ends. Not only is a truly flat illumination  not guaranteed with such a set-up, but the flat light enters the fibres at different angles from sky illumination and so the amount of light that actually enters each fibre depends on the fibre position within the field. Dome flats (a flat field produced by  reflecting light off a screen mounted on the dome) and sky flats are taken from time to time, but not every night. These two give an illumination of fibres that is much closer to that of stars. 

One flat field is created from all flap flats taken in one night for each arm and each plate. Since the fibres are repositioned roughly every hour, this averages out uneven illumination and produces evenly illuminated and consistent flats to use in the pipeline. See also Section \ref{sec:sky} for a description of the sky subtraction process, which is highly reliant on correct flat field reduction. Success in removing the sky emission lines is proof that averaging several flap flats produces a usable flat field image.

A flap flat field does not illuminate the whole CCD uniformly, so we do not correct images for a flat field response in a conventional way. Dome flats are not taken as part of a regular observing run due to time constraints and the twilight flats constitute the solar spectrum, which creates additional complications in the flat field correction. Due to simplicity we therefore use flap flats only anverywhere in the pipeline. We also do not use dark frames, as the HERMES CCDs are cooled to levels where dark current is negligible (measured dark current is less than 1 electron per hour). The first step in the reduction process is then bias correction. We use a median of all bias frames (typically there are 15 bias frames taken for each arm) for this, and we check that the bias levels are the same in the overscan of every image as in the combined bias image. If not, the overscan is used and subtracted instead of the combined bias frame.

Our science exposures suffer from three cosmetic aberrations: vertical streaks, degraded columns, and cosmic rays.

\begin{figure}
	\includegraphics[width=\columnwidth]{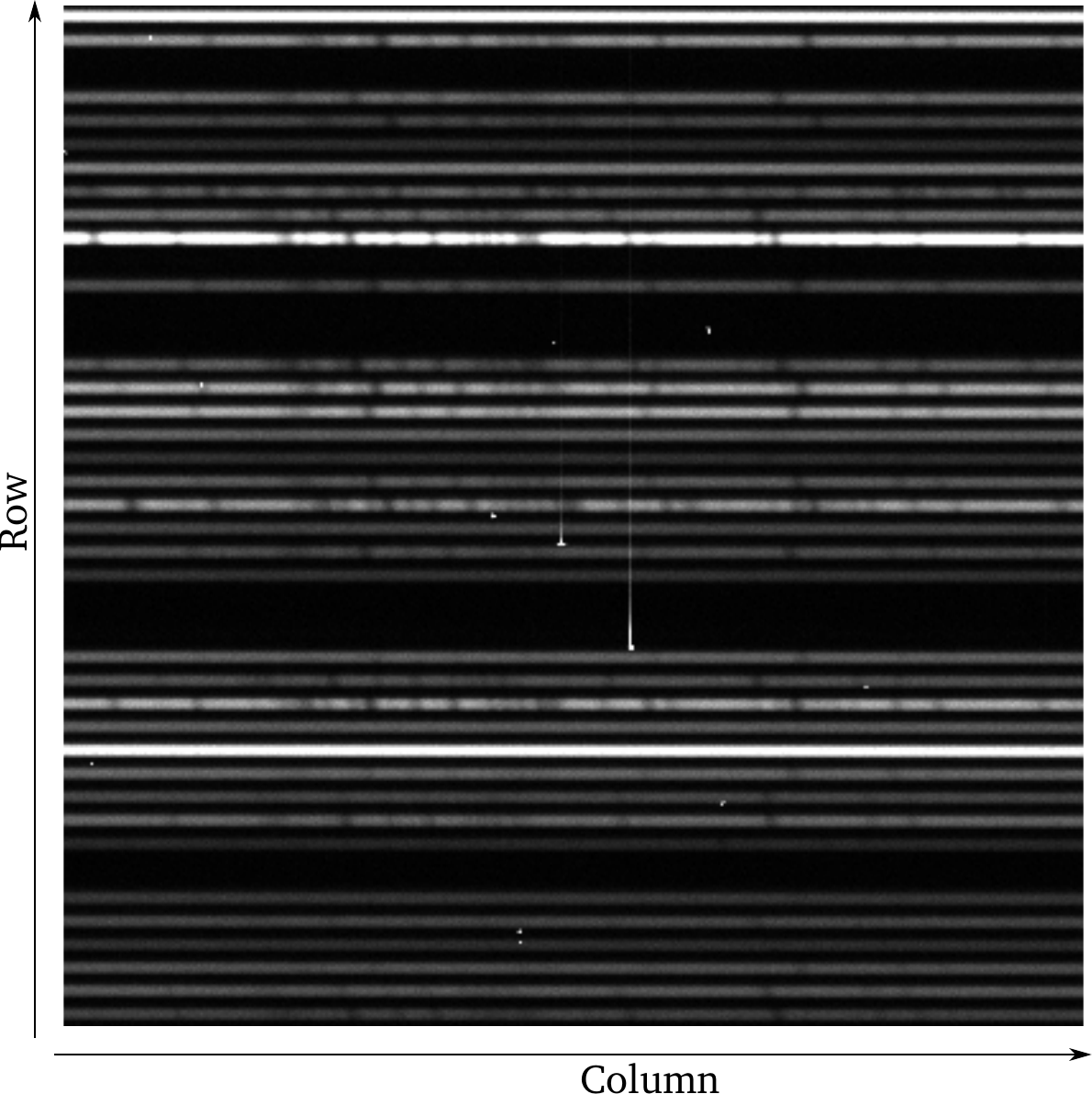}
    \caption{An example of two vertical streaks in a science exposure, made on February 8th 2015, run 39, green arm. Horizontal lines are traces of stellar spectra with clearly visible stellar absorption lines. Two streaks are visible in the centre. A few cosmic rays are also visible. Size of this image excerpt is $400 \times 400$ pixels.}
    \label{fig:strike}
\end{figure}

Vertical streaks are suspected to be very energetic particles originating from the HERMES spectrograph itself. The source has not been confirmed yet, though we suspect either one of the optical components or its coatings. There are around 20 such events detected per 20 minute exposure in the blue and green arms, only a few in the red arm, and none in the IR arm. The energetic events responsible for the streaks are visible in the IR arm but they do not produce the streak. A vertical strike is different from a traditional comic ray -- the strike saturates several pixels at the impact point and blooms into pixels in the same column (see Figure \ref{fig:strike}). The signal continuing from the blooming part becomes weaker with distance from the impact point and degrades to undetectable levels after 500 pixels in the most extreme cases and around 80 in a typical case. This means that a single vertical streak event impacts many spectra. Some vertical streaks are persistent between images and cannot be erased by a reasonable number of pre-flushes of the CCD. 

Fortunately a significant part of any strike is removed by the cosmic ray removal procedure. What is left is the centre of the streak (if it is large), and the weak tails of the streaks which do not have enough contrast to be detected by a cosmic ray removal algorithm.

Degraded columns are only present in the IR arm on the first ADC (analog-to-digital converter). There are 6 such columns (columns numbers 3262, 3270, 3271, 3308, 3324, and 3347) and the signal in them is replaced by linearly interpolated flux from nearest unaffected columns with the Iraf's \texttt{ccdproc} routine. They are all in the region of the spectrum uncorrupted by the telluric A band.

We are using a well-known and well-tested algorithm for detecting and removing cosmic rays called LaCosmic \citep{vandokkum01}. A Python version of the code is being used with minor modifications. The method described by \citet{vandokkum01} was designed to find and remove cosmic rays in images with no apparent patterns or symmetries, such as plain images of astronomical objects. However, images of spectra have a strong pattern in them (i.e. horizontal spectral traces). When a cosmic ray is detected, the flux in affected pixels can no longer be interpolated from the neighbouring pixels without considering this pattern. In the release of the code originally adopted by us, the affected pixels have flux interpolated from a box of $5\times5$ pixels in size. A median signal in this box is used to calculate the interpolated value in the pixel affected by a cosmic ray. We changed the box size to $7\times3$ pixels, with the longer side along the wavelength axis. This box size takes the obvious spectral pattern into account, and the interpolated value will be calculated from the pixels lying on the same spectral trace. Note that each spectral resolution element is around 4 pixels wide. The interpolated value is therefore calculated from the nearest two regions in the spectrum that can still be resolved, so the spectrum degradation in a pixel affected by a cosmic ray is minimal. This approach is not perfect, as the spectra also have a predictable pattern in the axis perpendicular to the spectral traces -- a distinct bell shape. Just changing the box shape, however, introduced a significant improvement.

In the blue and green arms, where we have strong vertical streaks, 0.073\% of pixels are detected to be affected by cosmic rays by LaCosmic in a 20 minute exposure. This drops to 0.062\% in the red and IR arms. We estimate that 99\% or more of the pixels affected by cosmic rays and vertical streaks are corrected. We are satisfied with the LaCosmic performance, so this is the only measure made to remove the cosmic rays. Even though we make 3 consecutive exposures of each field, we do not take their median at any stage. Median combination of, say normalized spectra, is often considered a good step to remove comic rays, but in our case it would fail to remove many vertical streaks because they persist through several exposures. We therefore find this step redundant, but still give the user an option to combine the individual spectra in any way possible.

\subsection{Tracing the spectra}
Tracing of the spectra is done on an averaged flat field. This means that a single tracing is used for a whole group of images (i.e. one night, one CCD, and one plate).

\begin{figure}
\includegraphics[width=\columnwidth]{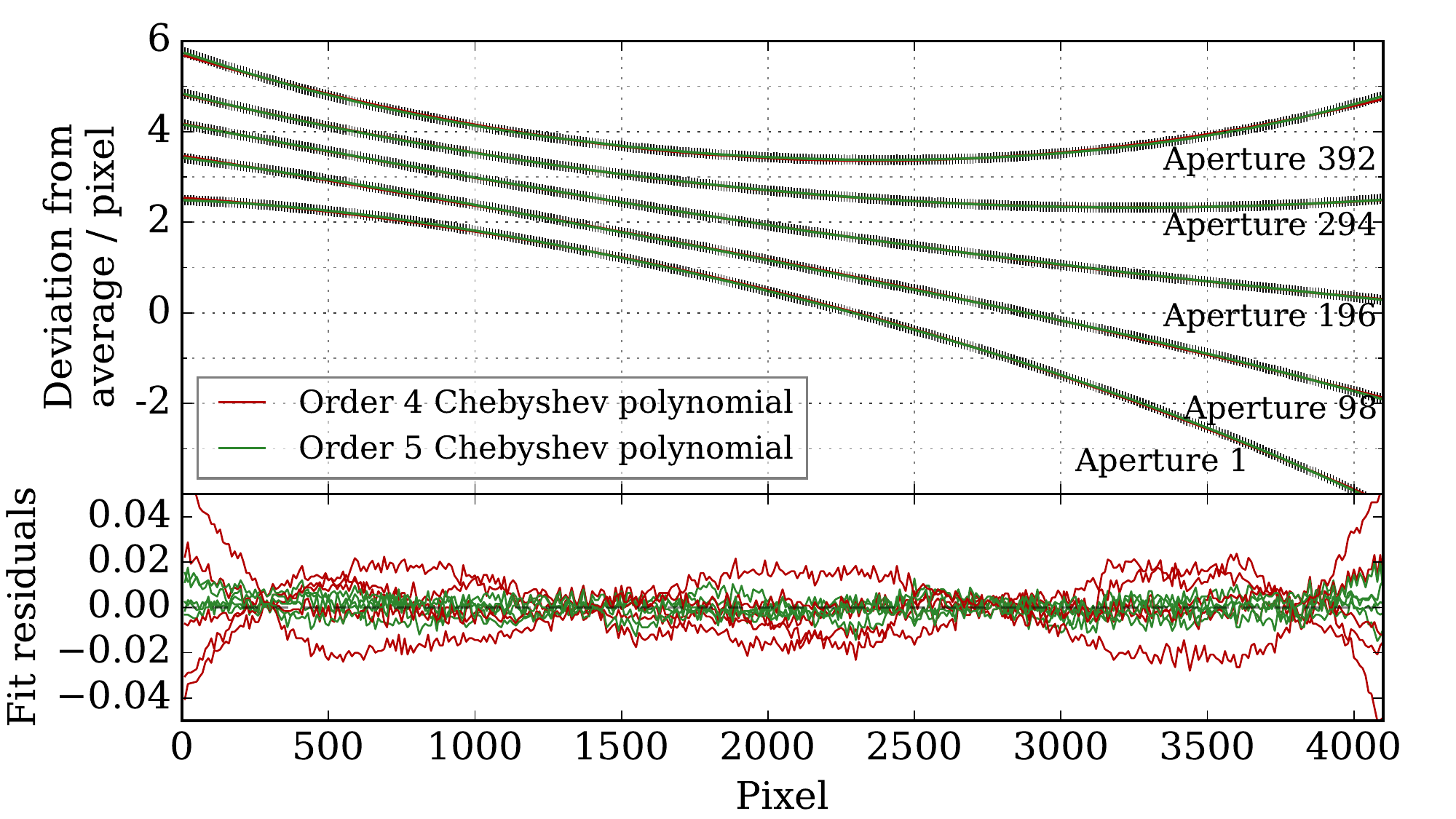}
\caption{Traces fitted in the apertures 1, 98, 196, 294, and 392 in a green arm flat field show the geometry of the spectral traces (top panel). Ploted traces are shifted for 1 pixel each in vertical direction. A Chebyshev polynomial of order 4 and 5 is fitted to each trace and the residuals are plotted in the bottom panel.}
\label{fig:traces}
\end{figure}

392 extraction apertures must be positioned on the spectral traces in each image. A key (as described above) is used to give the pipeline an approximate position for each aperture in one reference column (see Section \ref{sec:nonauto}) of the image. After the centre of the aperture is calculated at the reference column it is propagated along the spectrum trace. Iraf's \texttt{center1d} algorithm is used to find the aperture centeres along the trace. 10 consecutive columns are averaged before the center of the trace is found, so the algorithm is run on higher SNR data and fewer times for each trace. The shape of the trace is represented by a 5th order Chebyshev polynomial. The traces are almost straight, well behaved and follow one row of pixels with a deviation of 3 pixels at most. The width of the aperture is fixed at 6 pixels and is centred on the fitted trace. Pixelization is taken care of by Iraf's \texttt{apfit} and \texttt{fit1d} algorithms.

\subsection{Correction of optical aberrations}

\begin{figure*}
\setlength{\tabcolsep}{1pt}
\begin{tabular}{ccc}
\multirow{4}{*}{\rotatebox{90}{$\xrightarrow{\makebox[1.5cm]{Sub-apertures}}$}}&7&\multirow{4}{*}{\includegraphics[width=0.88\textwidth]{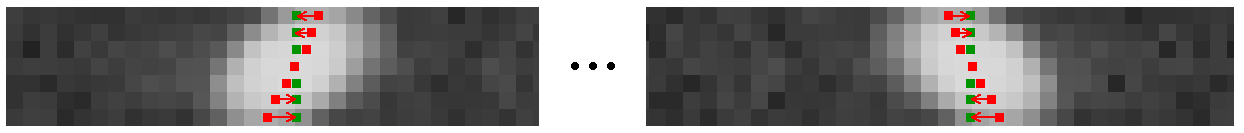}}\\[1mm]
&5&\\[1mm]
&3&\\[1mm]
&1&\\[0.2cm]
\multirow{4}{*}{\rotatebox{90}{$\xrightarrow{\makebox[1.5cm]{Sub-apertures}}$}}&7&\multirow{4}{*}{\includegraphics[width=0.88\textwidth]{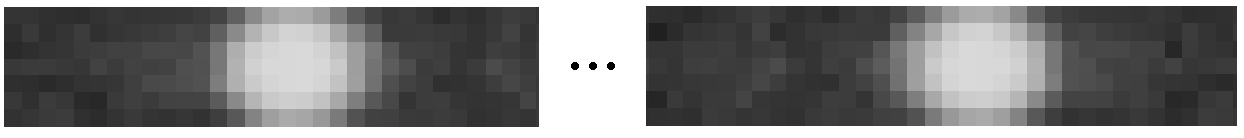}}\\[1mm]
&5&\\[1mm]
&3&\\[1mm]
&1&\\[0.2cm]
&&$\xrightarrow{\makebox[0.85\textwidth]{ Wavelength }}$
\end{tabular}
\caption{Example of two arc lines in the corners of a red arm image, before (top) and after (bottom) the geometric correction. Left image shows the shape of an arc line in the top left corner and right image shows it in the top right corner of an image. Notice that optical aberrations distort the two arc lines differently. The aperture is divided into 7 sub-apertures, seen here as rows of pixels. Sub-apertures are equal division of an aperture and are not centred on pixels but run parallel to the spectrum trace. The calculated centre of the line in each sub-aperture is marked with a red dot and the centre of the line as measured over the whole aperture is marked in green. A transformation has to be found that moves the red points onto the green ones, as illustrated by red arrows. The resolution in the bottom panel is improved by $\sim15$\% over the resolution in the top panel.}
\label{fig:aberr}
\end{figure*}

HERMES is designed to produce almost linear, equally spaced, two dimensional spectral traces. This is achieved to a high degree. Optical aberrations are most prominent in the corners of the CCD, where the point spread function (PSF) takes an obvious elliptical shape. A similar distortion is observed in all four arms. Because the PSF of the HERMES spectrograph is not predictable enough to account for any time changes and changes due to focus, temperature, etc., we choose to correct this aberration empirically, by using arc lines to measure the tilt of the elongated PSF and straightening the spectra to narrow the PSF in the wavelength direction (see Figure \ref{fig:aberr} for an example of a ''tilted'' PSF and the desired correction). This way we can improve the spectral resolution in the detector corners. We conclude that the PSF is adequately sampled by arc lines for this operation. We do this process for every aperture individually.

\begin{figure}
\centering
\includegraphics[width=0.8\columnwidth]{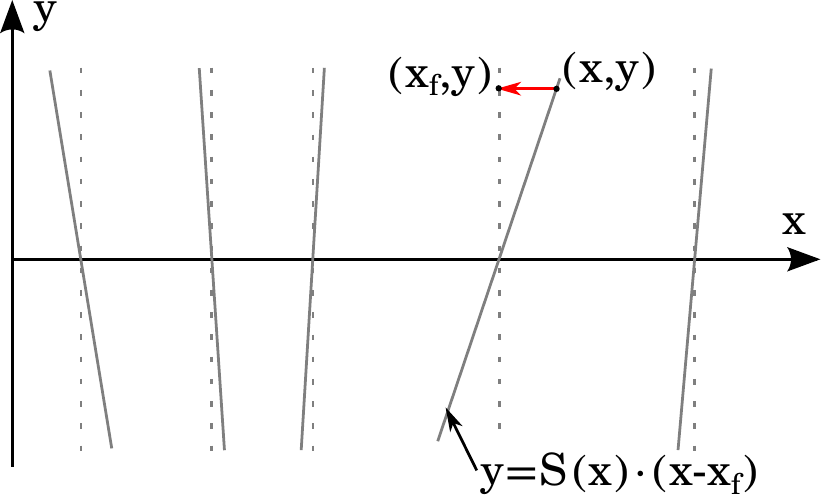}
\caption{An illustration of the problem where a varying shear has to be fitted. The $x$ coordinate runs along the wavelength axis, and the $y$ axis along the sub-apertures (with the middle sub-aperture having the value $y=0$). Tilted grey lines represent the shear measured from arc lines at different wavelengths, and dashed lines show what the image should look like after the geometric transformation is applied. The equation of each solid grey line is $y=\mathrm{S}(x)\cdot(x-x_f)$, where the shear $\mathrm{S}(x)$ varies from line to line.}
\label{fig:draw}
\end{figure}

Each aperture is divided into 7 parallel and equal-width sub-apertures and a spectrum is extracted from every sub-aperture. This is performed by Iraf's \texttt{apall} routine with option \texttt{nsubaps} set to 7. Arc lines are identified in each sub-aperture, but no wavelength calibration is done. Only the positions of centres of the lines in the pixel space are recorded. They are calculated by Iraf's \texttt{center1d} algorithm. It is then assumed that the centre of an arc line should have the same coordinate in every sub-aperture, and this gives us a condition for a transformation. For every arc line, we want to transform the coordinates of the line centres in each sub-aperture to match the global coordinate, as measured if there was only one aperture:
\begin{equation}
\begin{aligned}
(x_1,1)\rightarrow(x_{f},1)\\
(x_2,2)\rightarrow(x_{f},2)\\
\vdots\qquad\qquad\\
(x_7,7)\rightarrow(x_{f},7)\\
\end{aligned}
\label{eq:1}
\end{equation}
where $x$ is the coordinate in pixels along the wavelength axis, and $x_f$ is the global coordinate of each arc line. The second coordinate component is just the sub-aperture number (see Figures \ref{fig:aberr} and \ref{fig:draw} for an illustration).

\begin{figure}
\centering
\includegraphics[width=0.95\columnwidth]{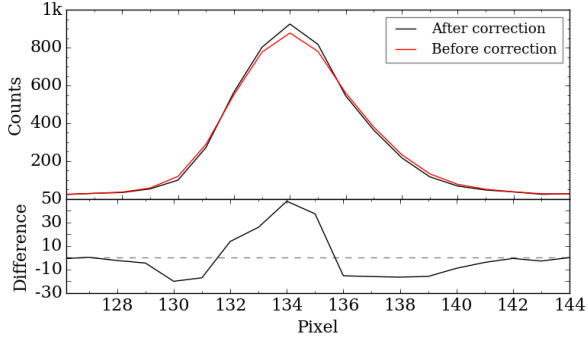}
\caption{Profiles of the arc line in the top-left part of the red image before (red) and after (black) the geometric correction is applied (top panel). This arc line is also displayed in the left two panels of Figure~\ref{fig:aberr}. The difference between the two profiles is plotted in the bottom panel. The spectrum has already been colapsed into one dimension in this plot.}
\label{fig:arc_geom}
\end{figure}

With $\sim40$ arc lines per spectrum and 7 data points coming from the above transformation condition for each line we have hundreds of data points to constrain the following transformation. The transformation adopted is a linear skewness that can change along the wavelength axis according to a 3rd order polynomial. Following the rule in Equation \ref{eq:1}, a transformation is written for each pixel:
\begin{equation}
y=\mathrm{S}(x)\cdot(x-x_f)
\end{equation}
where $\mathrm{S}(x)=Ax^3+Bx^2+Cx+D$ is a 3rd order polynomial describing how the shear is changing along the $x$ axis. What we are interested in is the $x_f(x,y)$, the shift we have to apply to each pixel, which can be calculated from the formula
\begin{equation}
x_f(x,y)=x-\frac{y}{Ax^3+Bx^2+Cx+D}
\label{eq:xf}
\end{equation}
after the coefficients $A$ to $D$ are fitted to measured points by minimizing the difference between the $x_f$ from Equation \ref{eq:xf} and the measured shifts for each line (as shown in Equation \ref{eq:1}). Five iterations of the fitting process are done, with outliers above 2.5 sigma rejected in each iteration. This was the simplest transformation that produced a significant and satisfactory correction. In practice the transformation is calculated by Iraf's \texttt{geomap} and performed by \texttt{geotran} routines, because they are simple to integrate with routines used to find the arc lines and create the subapertures.

The transformation is calculated for each aperture. The optics are stable enough that the transformation need only be calculated once per night and applied to all images. While a single transformation could, in principle, be used for repetitive nights, we calculate a transformation for each night as we reduce each night independently. Both arc and science images are corrected following the same transformation, which produces a significant improvement in extracted resolution of $\sim15$\% in the detector corners.

\subsection{Scattered light}
Scattered light is an anomalous, large scale excess signal originating from the scattering of light in the optical components, the detector itself and, in a very small part, the air in the spectrograph. While HERMES has exceptionally low scattering for a fibre spectrograph, it is still detectable and must be corrected.

\begin{figure}
	\includegraphics[width=\columnwidth]{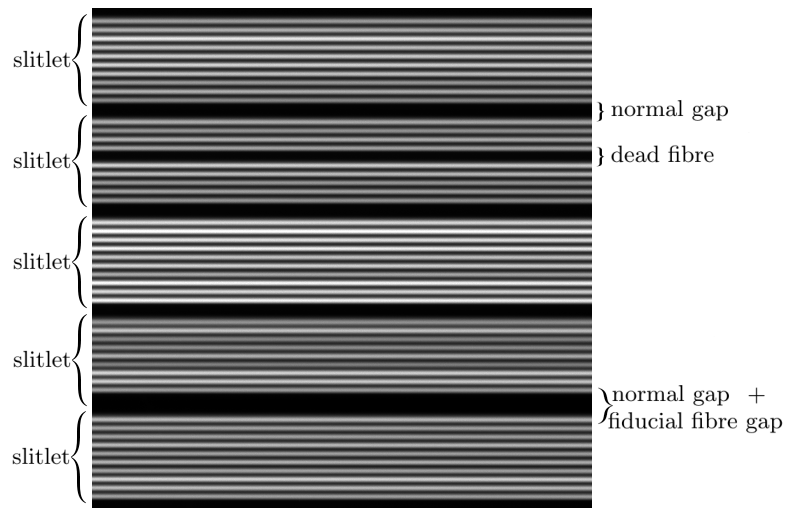}
    \caption{Slitlets and gaps between slitlets presented in a flat field image. Size of this image excerpt is $500 \times 500$ pixels.}
    \label{fig:scattered}
\end{figure}

As described earlier, fibres are organised into slitlets, with a significant gap between them. These gaps allow us to measure the amount of scattered light in individual images (see Figure \ref{fig:scattered}). There are also 8 gaps that are almost twice the width between the slitlets, because a fibre that should be next to a gap is a fiducial fibre (used for guiding) and is therefore not connected to the spectrograph. No signal should be present in such gaps, so all the signal we detect there can be attributed to  scattered light and fibre cross-talk. While there are also gaps due to dead fibres, these are narrow and there is no guarantee that the fibre has exactly zero throughput.

Apertures are always 6~pixels wide and centred on spectrum traces. We define gaps as regions at least 0.7~pixel away from the edge of any aperture. A normal gap between slitlets is 22.5~pixels wide, measured as the distance betwen trace centres on each side of the gap. Taking the width of the apertures and the 0.7~pixel buffer into the account, the usable part of the gap is 15.1~pixels wide or more, if there is a fiducial fibre in the gap. As it will become evident in Section \ref{sec:fct}, a large part of any gap must be free of any signal coming from the wings of the spectral traces, so it is only affected by the scattered light.

We use the Iraf's \texttt{apscatter} routine to measure, fit, and subtract the scattered light. First the cross-sections of gaps perpendicular to the dispersion axis are constructed and the signal in 70 consecutive columns is averaged. This averaging produces sufficient SNR to select the parts of the gaps where signal is lowest. This is achieved by a highly asymmetric sigma clipping algorithm (low rejection limit is 4 sigma and high rejection limit is 1.1 sigma). These regions are assumed to sample the scattered light. The rest of the signal in the gaps will be attributed to fibre cross-talk in a later step. The scattered light is then measured and fit along the dispersion axis, but only in the regions defined in the previous step. A two dimensional 6th degree Chebyshev polynomial is used to represent the scattered light in both directions and is then subtracted from the entire image. 

\begin{figure}
	\includegraphics[width=\columnwidth]{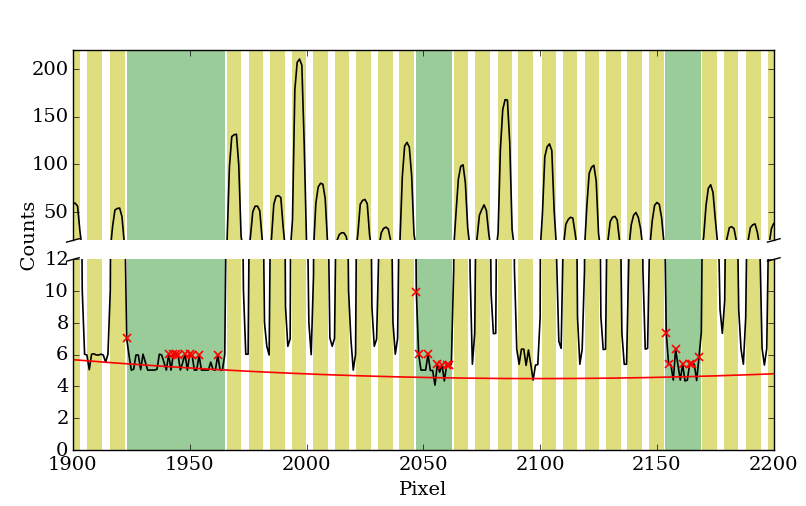}
    \caption{Fitted scattered light. The black line is the median of a spectrum in 70 columns. Yellow regions are extraction apertures, and green regions are gaps. The fitted level of scattered light is the red line. Crosses show points in the gaps that were rejected by the sigma clipping algorithm. Only a small portion of the whole image is shown here.}
    \label{fig:scattered2}
\end{figure}

Figure \ref{fig:scattered2} shows one of the cross-sections with a fitted polynomial and rejected points. 59 such cross-sections are made over the whole image.

\subsection{Fibre cross-talk}
\label{sec:fct}
\begin{figure}
	\includegraphics[width=\columnwidth]{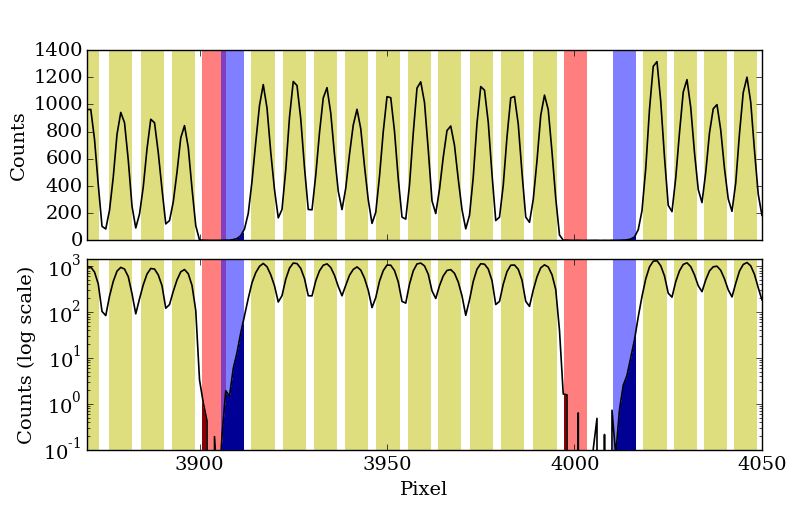}
    \caption{Cross-section of a flat field showing positions of the phantom apertures in the process of the cross-talk measurement. Yellow regions are regular apertures, red are succeeding phantom apertures and blue are preceding phantom apertures, both placed in a gap. The flux that counts as a cross-talk is shaded in darker colours. The figure is presented in linear and logarithmic scales to give the right impression on the fibre cross-talk values. We show a corner of the flat field, where the fibre cross-talk is greatest.}
    \label{fig:ct1}
\end{figure}

Fibre cross-talk\footnote{We use the term fibre cross-section for what should be technically called a cross-talk between apertures. Since the effect on the final one-dimensional spectrum is the same for both kinds of cross-sections, we prefer to use a more common term fibre cross-section.} is the fraction of the light originating in one fibre and being detected in the region of the CCD (in the aperture) where a neighbouring fibre contributes its light.

Fibre cross-talk is measured by using the slitlet gaps again. Because the fibre cross-talk varies from fibre to fibre, it is impossible to measure it for each fibre individually. However the variations are small and predictable, so the fibre cross-talk can be measured for each slitlet and interpolated to individual fibres. The amount of fibre cross-talk is independent of the spectrum in each fibre and can be measured in a flat field image. Two fibre cross-talks must be measured separately: one for the leak from the preceding and one for the leak from succeeding fibres. If a fibre is next to a gap, we assumed there is no cross-talk from that side.

\begin{figure*}
\includegraphics[width=\columnwidth]{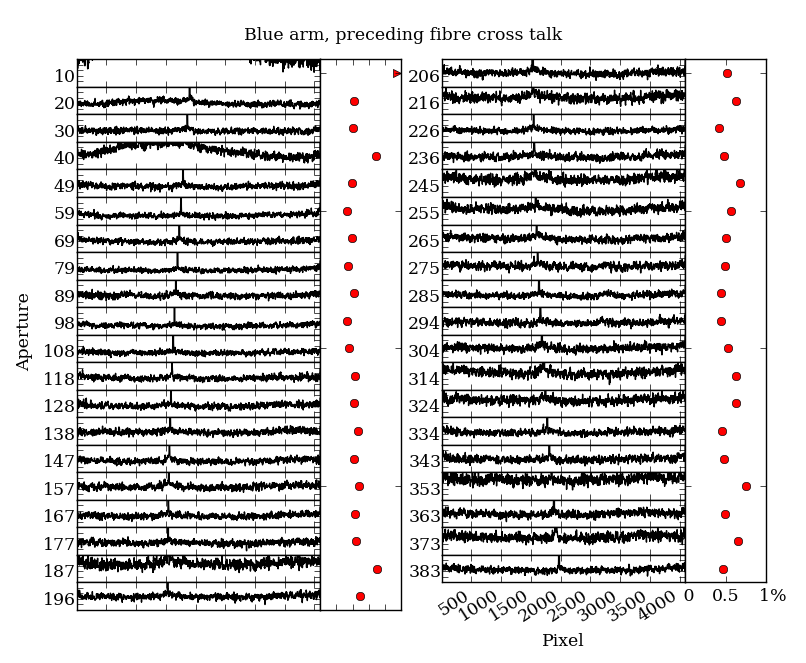}\includegraphics[width=\columnwidth]{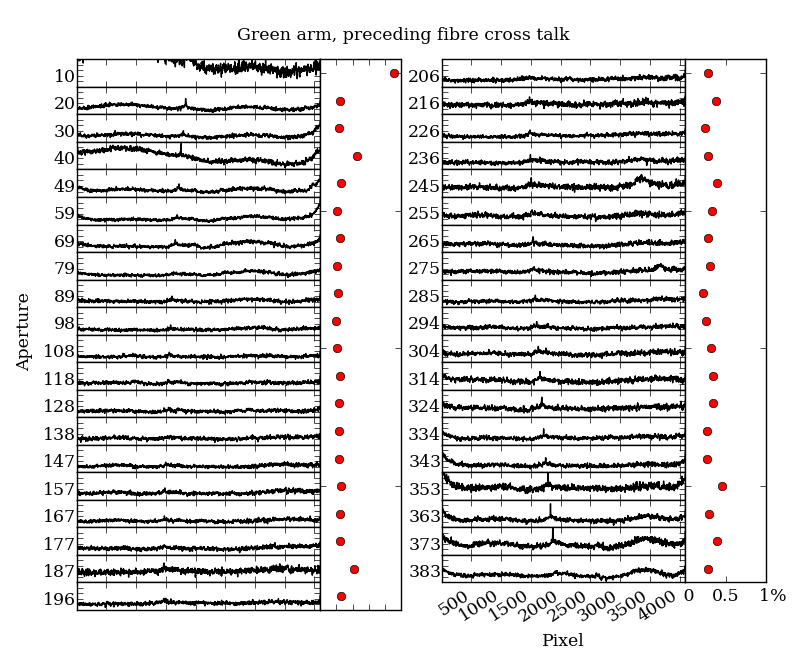}\\
\includegraphics[width=\columnwidth]{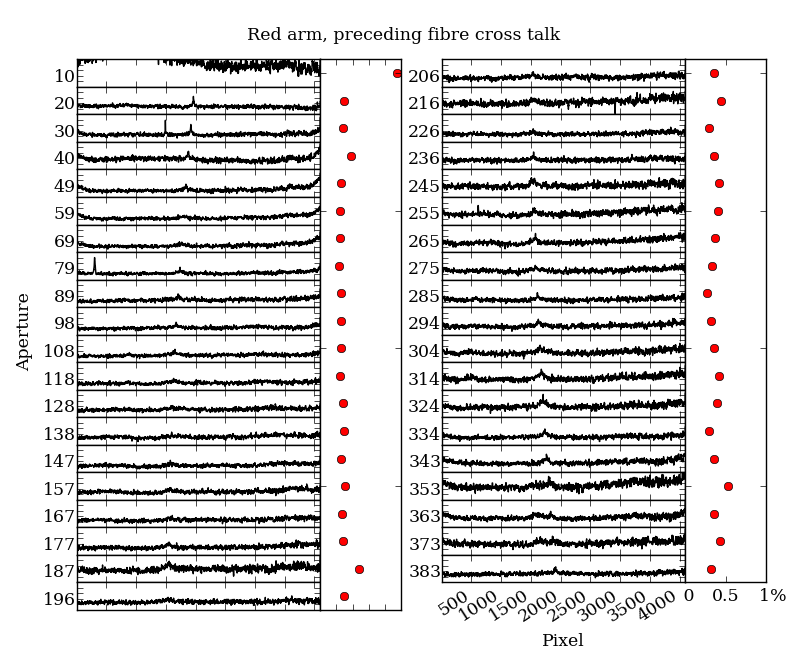}\includegraphics[width=\columnwidth]{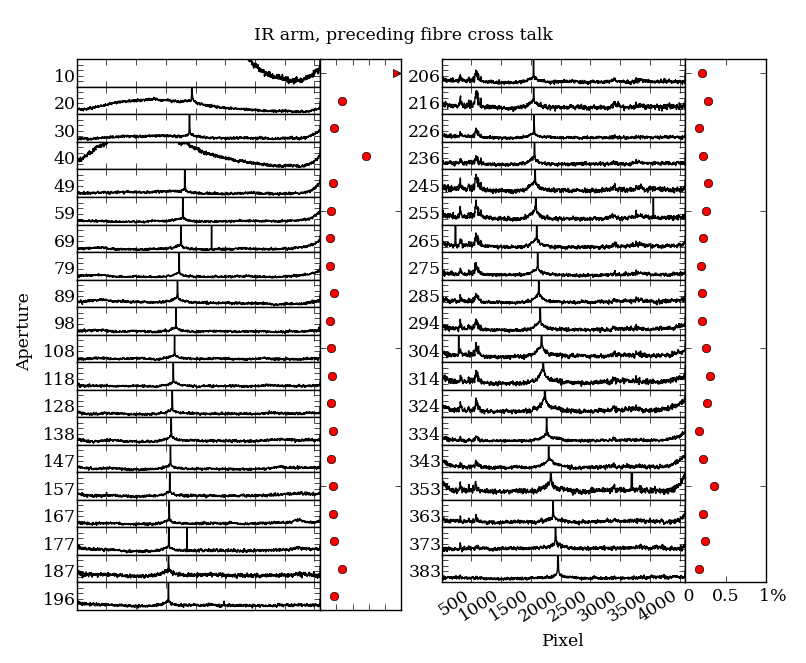}
\caption{Variation of the preceding fibre cross-talk as it depends on the wavelength (here in pixel scale) and fibre (here given as the aperture number). There are 39 gaps in which we can measure the fibre cross-talk, hence 39 panels. Aperture number is the number of the aperture next to the gap in which the fibre cross-talk was actually measured. Every panel has a vertical scale between 0 and 1\% amounth of fibre cross-talk. The adjacent panel with red dots shows the average fibre cross-talk for each panel on the left from the dot. 39 panels are split into two columns. A spike in the middle of each panel is a ghost from the 0 order spectrum. It is removed from science frames, but not from the flat field used here.}
\label{fig:cross2}
\end{figure*}

\begin{figure*}
\includegraphics[width=\columnwidth]{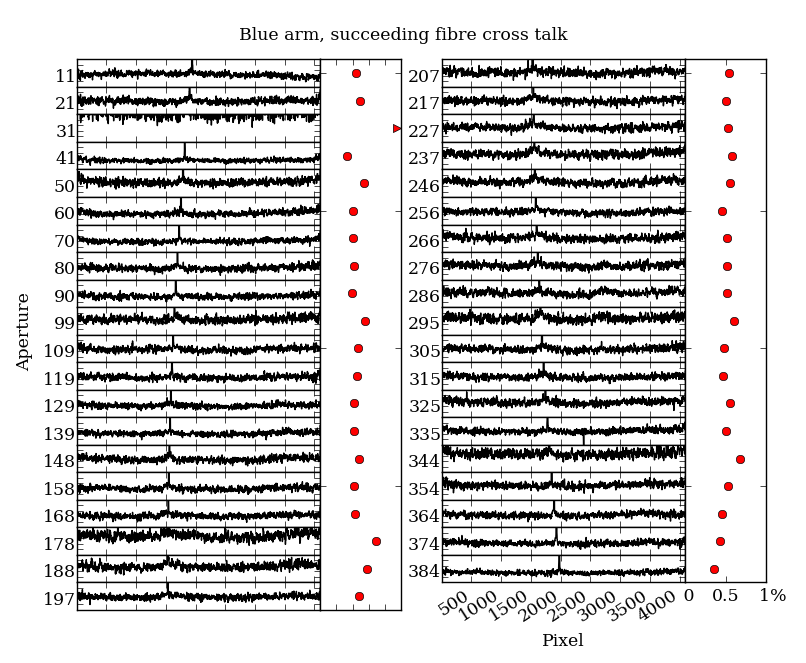}\includegraphics[width=\columnwidth]{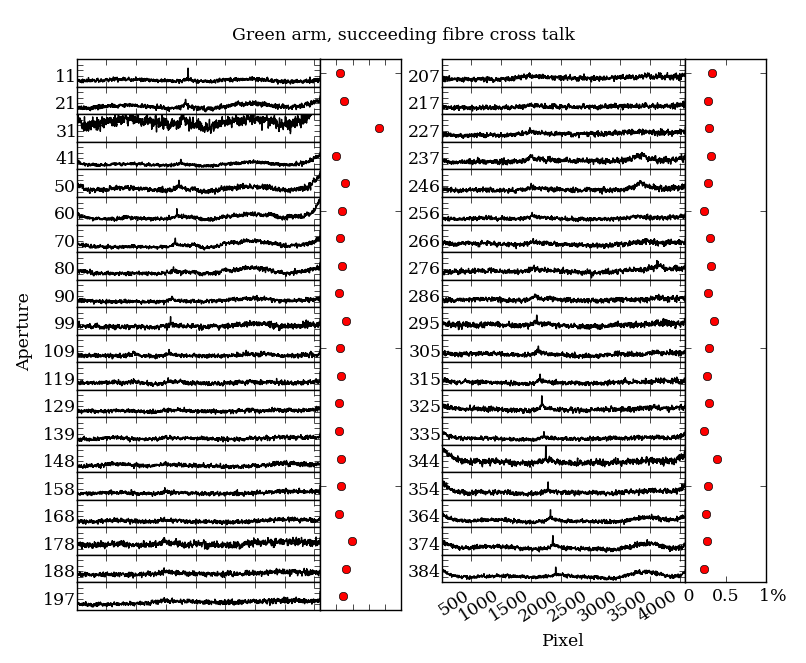}\\
\includegraphics[width=\columnwidth]{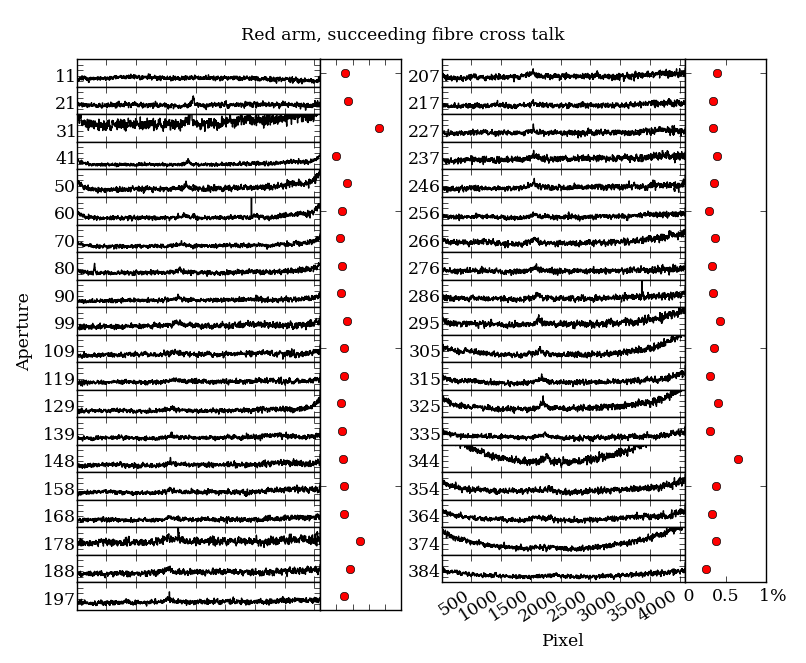}\includegraphics[width=\columnwidth]{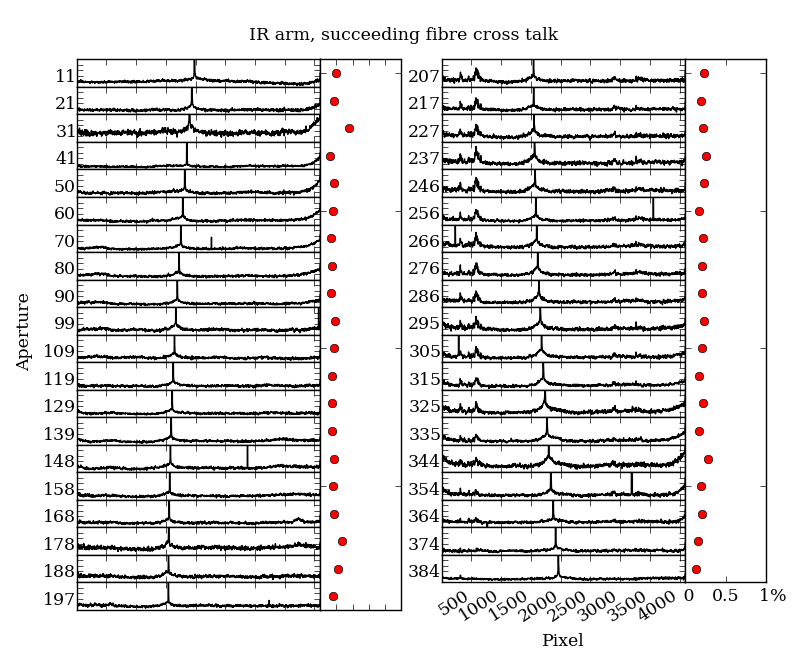}
\caption{Same as Figure \ref{fig:cross3} but for the succeeding fibre cross-talk.}
\label{fig:cross3}
\end{figure*}

For each slitlet we add two so called phantom apertures, one before the slitlet and one after the slitlet (as illustrated in Figure \ref{fig:ct1}), and extract the spectra in all the apertures, including the phantom ones. Because the flux in phantom apertures should be zero, the actual measured flux is just the fibre cross-talk from the neighbouring aperture. Dividing this by the flux in the proper aperture, we get the amount of fibre cross-talk for this slitlet. Each slitlet has a preceding and a succeeding phantom aperture, and we measured both cross-talk values. There are 39 gaps in every image, however, some are rejected from this calibration because they include ghosts, as illustrated in Figures \ref{fig:cross2} and \ref{fig:cross3}. The amount of cross-talk is interpolated with a 2-dimensional 6th-order polynomial. This value is then used to subtract the fibre cross-talk from every spectrum. The actual spectrum from the neighbouring fibre to be subtracted is smoothed beforehand with a 5 pixel median box filter to prevent any residual cosmic rays or vertical streaks from propagating into neighbouring spectra.

\begin{figure*}
\begin{tabular}{cccc}
\multirow{8}{*}{\rotatebox{90}{$\xrightarrow{\makebox[21.6cm]{ Aperture number }}$}} & Blue arm, preceding fibre cross-talk & Blue arm, succeeding fibre cross-talk & \\
& \includegraphics[width=0.65\columnwidth, height=0.6\columnwidth]{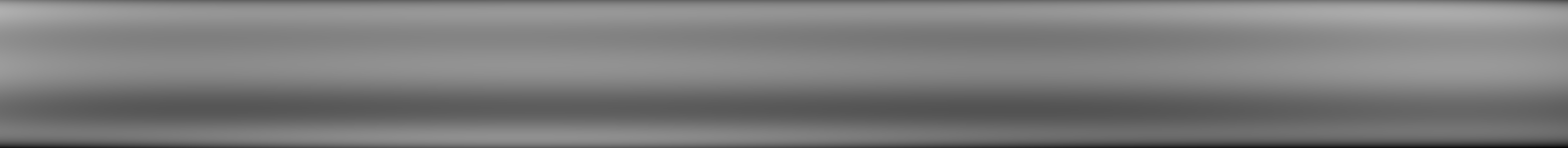} & \includegraphics[width=0.65\columnwidth, height=0.62\columnwidth]{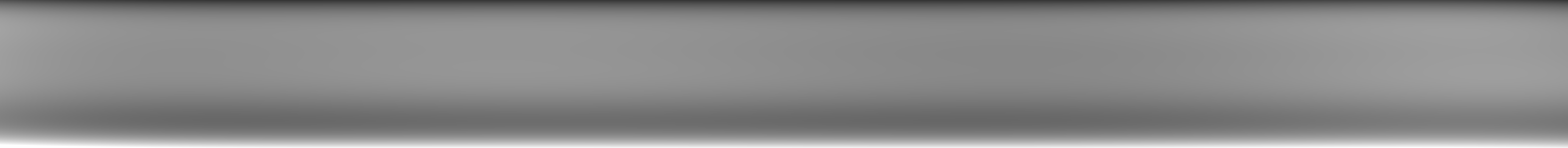} & \multirow{7}{*}{\includegraphics[width=1cm]{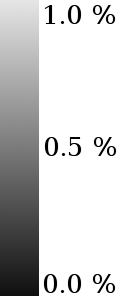}}\\
& Green arm, preceding fibre cross-talk & Green arm, succeeding fibre cross-talk & \\
& \includegraphics[width=0.65\columnwidth, height=0.62\columnwidth]{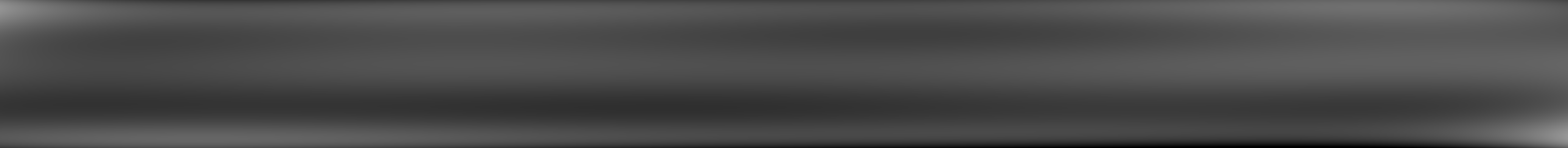} & \includegraphics[width=0.65\columnwidth, height=0.62\columnwidth]{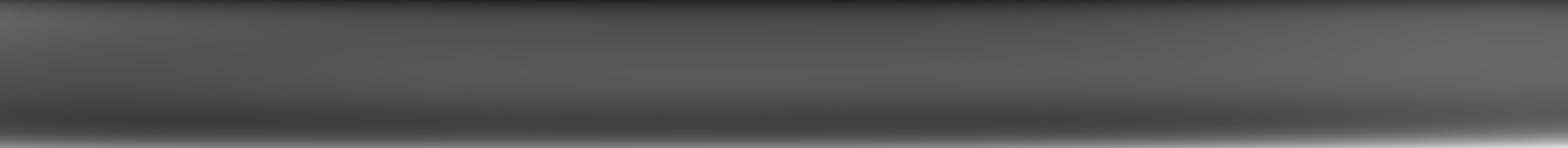} & \\
& Red arm, preceding fibre cross-talk & Red arm, succeeding fibre cross-talk & \\
& \includegraphics[width=0.65\columnwidth, height=0.62\columnwidth]{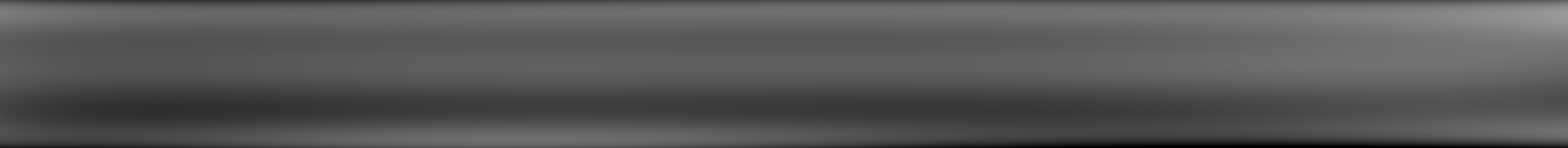} & \includegraphics[width=0.65\columnwidth, height=0.62\columnwidth]{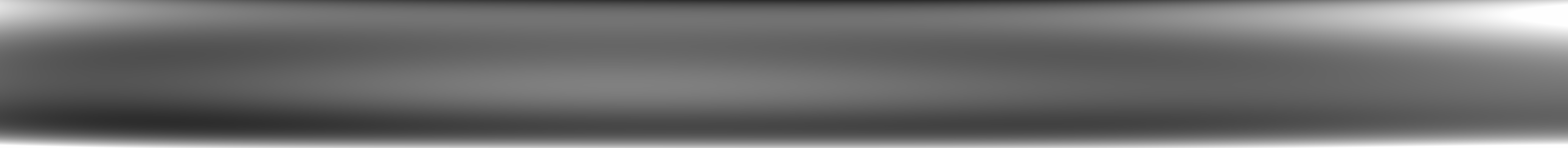} & \\
& IR arm, preceding fibre cross-talk & IR arm, succeeding fibre cross-talk & \\
& \includegraphics[width=0.65\columnwidth, height=0.62\columnwidth]{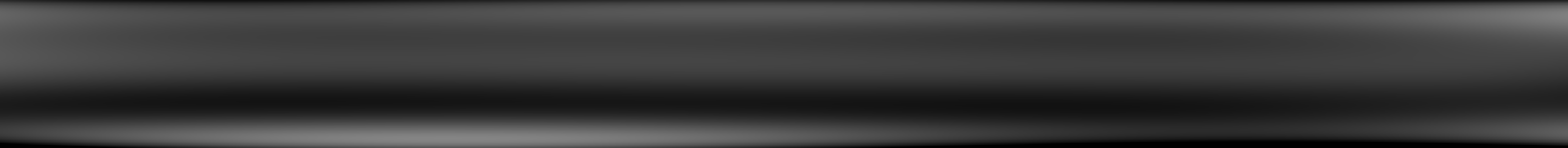} & \includegraphics[width=0.65\columnwidth, height=0.62\columnwidth]{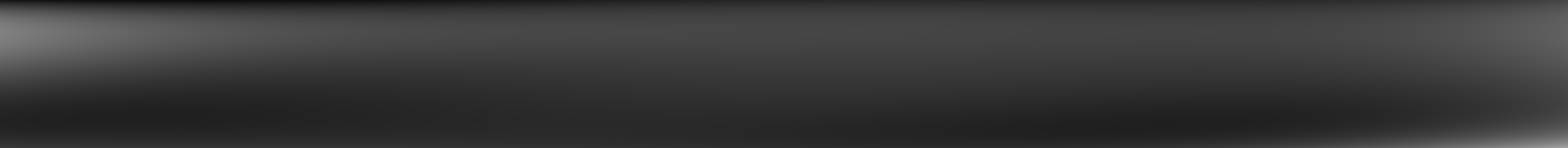} & \\
& \multicolumn{2}{c}{$\xrightarrow{\makebox[0.73\textwidth]{ Wavelength }}$} & \\
\end{tabular}
\caption{Fitted and smoothed fibre cross-talk for every pixel and for every fibre. There are 4094 pixels per fibre and 392 apertures (8 fiducial fibres are not shown). The gray-scale shows the amount of fibre cross-talk and is normalized to 0 to 1\% range.}
\label{fig:maps}
\end{figure*}

A formula to correct for the fibre cross-talk is then:
\begin{equation}
\begin{aligned}
\begin{array}{rl}
S_{F}^{corr}(x)=S_{F}(x)&-\left[S_{F-1}(x)M_{p}(x,F)\right]\delta(F,F-1)\\&-\left[S_{F+1}(x)M_{s}(x,F)\right]\delta(F,F+1)\\[0.25cm]
\end{array}\\[0.4cm]
\mathrm{where}\qquad\qquad\qquad\qquad\\[0.4cm]
\delta(F_1,F_2)= 
\begin{cases}
    0, &\text{if there is a gap between $F_1$ and $F_2$} \\
    1,              & \text{otherwise}
\end{cases}
\end{aligned}
\end{equation}

\noindent where $S_{F}$ is the spectrum in fibre number $F$, $x$ is a coordinate running along pixels in the wavelength direction, and $M_p$ and $M_s$ are preceding and succeeding fibre cross-talk maps, like in Figure \ref{fig:maps}.

The amount of fibre cross-talk depends strongly on the aperture width, which we have chosen to fix at 6~pixels. This means that some of the signal that does not fall into the aperture is lost, but at the same time the fibre cross-talk is kept low. An aperture width one pixel larger, for example, would mean that the fibre cross-talk is of the same order of magnitude as the signal gained by a wider aperture. 

Fibre cross-talk needs only be subtracted once. Further iterations are not needed because only stars in a two magnitudes range are observed at the same time. Any second-order fibre-cross talk is therefore undetectable.

The fibre cross-talk measurements demonstrate that the cross-talk changes smoothly from gap to gap, so we can safely assume that it is a result of spectrograph optics and should also change smoothly from fibre to fibre. This means that measuring the fibre cross-talk in the gaps alone is sufficient.

\subsection{Wavelength calibration}
\label{sec:wav}
The wavelength scale is calibrated using spectra from a Thorium-Xenon (ThXe) arc lamp. One arc spectrum is taken before or after each set of science exposures. It is used as a calibration only for the science exposures observed with that fibre configuration. 

\subsubsection{The {ThXe} lamp and line identification}
\label{sec:wavs}
The dominant lines in our lamp belong to Xe, which is not widely used in visual or near-infrared spectroscopy at wavelengths less than 8000~{\AA}. All the literature linelists \citep{hansen1987,palmer1983,meggers1933,lovis2007,redman2014,humphreys1939,meggers1934, zalubas1976,ahmed1998,zalubas1974} lacked the desired precision and were found to often be inconsistent with each other. We also see lines in our arc spectra that are not present in the literature linelists. As  Figure \ref{fig:arc_spectra} shows, we need to use as many of the available lines as possible for calibration. Th lines are generally much weaker than Xe lines, but do have well known wavelengths. However there are not enough Th lines to ensure a good wavelength calibration without using the Xe lines as well. The most problematic regions are in the green, red and IR arms where Th lines are either very weak, or totally absent.

To assemble a reliable line list, we made a set of special arc exposures with ThXe and Thorium Argon (ThAr) lamps. ThXe and Thorium-Argon(ThAr) lamps were turned on at the same time, so both sets of lines are exposed on the same image, removing any concerns about stability and time variation during the exposure. 360~s exposures were made for blue, green and red arms, which is double the time we use for regular arc exposures. In the IR arm we took a set of 120 second long exposures, otherwise some Ar lines would saturate and could potentially damage the CCD. 

The ThAr+ThXe arcs were wavelength calibrated against the ThAr line list provided in Iraf's \texttt{linelist/thar.dat} file (wavelengths for this line list are taken from \citet{palmer1983}). To assemble the ThXe line list, the wavelengths of Th and Xe lines were measured in the calibrated spectra. We inspected each line by eye and excluded lines that featured peculiar profiles such as strong asymmetry due to blended lines and double peaks. We also omitted weak lines that are lost in the noise in regular arc exposures, but just visible in longer exposures of ThAr+ThXe. 

The initial wavelength calibrations of the ThAr+ThXe lamp spectra done on Th and Ar lines have a root mean square (rms) residual of 0.0099~{\AA}, 0.0054~{\AA}, 0.0079~{\AA}, and 0.0129~{\AA} in the blue, green, red, and IR arms (respectively).

\begin{figure*}
\includegraphics[width=\columnwidth]{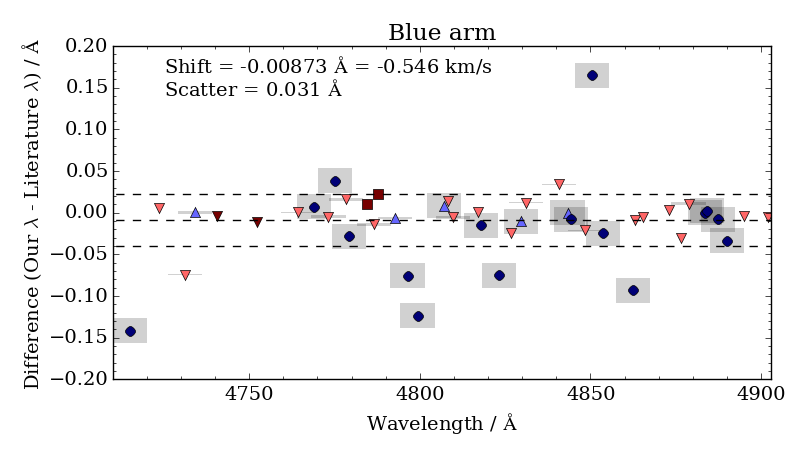}\includegraphics[width=\columnwidth]{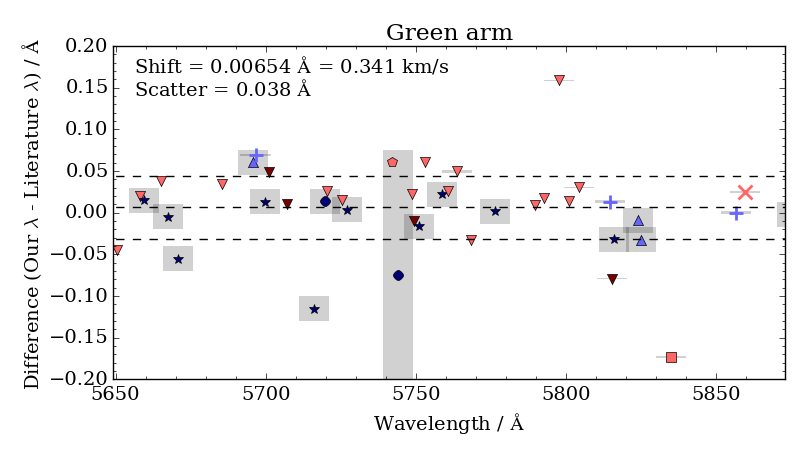}\\
\includegraphics[width=\columnwidth]{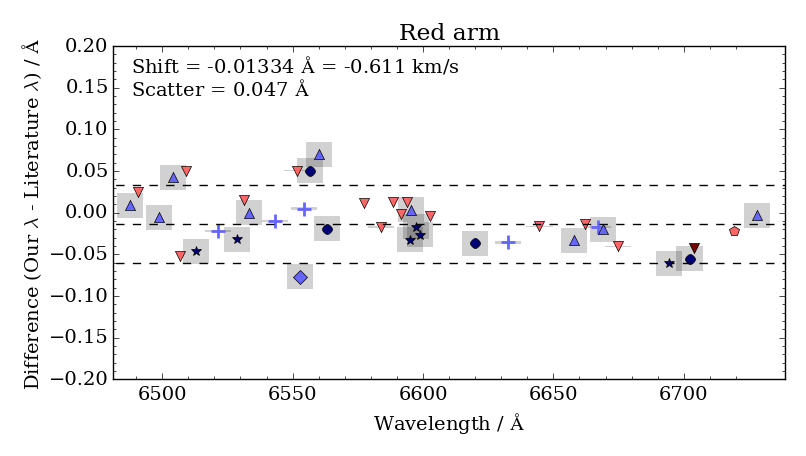}\includegraphics[width=\columnwidth]{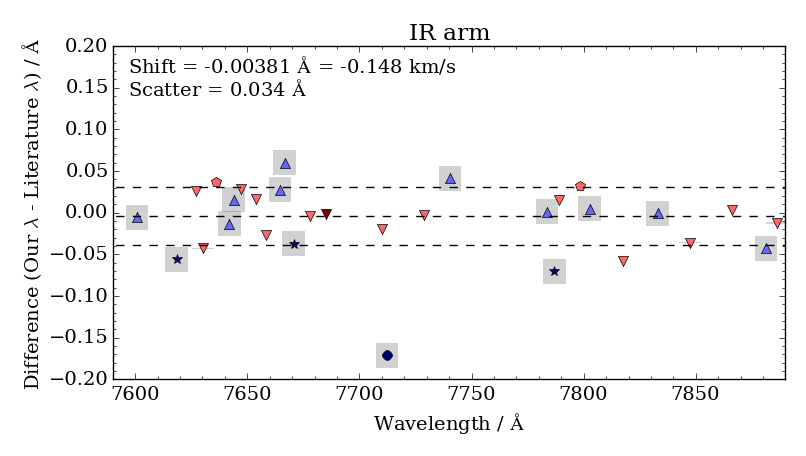}\\
\caption{Difference between the wavelengths in our line lists and wavelengths of the same lines in the literature. Dark red symbols represent Th II lines, bright red Th I, dark blue Xe II and light blue Xe I lines. Shape of the symbol marks the literature source ($\bullet$: \citet{hansen1987}, $\blacktriangledown$: \citet{palmer1983}, $\blacktriangle$: \citet{meggers1933}, $\blacksquare$: \citet{lovis2007}, $\pentagon$: \citet{redman2014}, $\bigstar$: \citet{humphreys1939}, $+$: \citet{meggers1934}, $\times$: \citet{zalubas1976}, $\blacklozenge$: \citet{ahmed1998}, $\hexagon$: \citet{zalubas1974} or respectively \textit{a} to \textit{k} in Table \ref{tab:linelist}). Gray rectangles show the precision of the wavelength given in the literature. The largest rectangle in the green arm plot means that the wavelength is given with 1 decimal place, next size is 2 decimal places. Average shift and dispersion are plotted with dashed horizontal lines. Lines that we did not find in the literature and have only the nearest line given in the table \ref{tab:linelist} are excluded from this calculation.}
\label{fig:linelists2}
\end{figure*}

This two-step process might introduce small systematic errors in the final measured wavelengths. Figure \ref{fig:linelists2} shows the discrepancies between our line lists and literature values. The large scatter is primarily due to imprecisions in the line lists from the literature, where mostly Xe lines have wavelengths measured from low-resolution spectra with a very limited accuracy. The shift is purely statistical. A similar amount of shift is seen when comparing lines from different literature sources among themselves, so having a similar shift against our line list is expected. We further demonstrate that the shift is coincidental in Section \ref{sec:rv} and Figure \ref{fig:rv_dif}. Our final line lists are collected in Appendix \ref{sec:linelist2}.

\subsubsection{Wavelength solution}
The apertures derived from the flat field are used to extract science and arc spectra with exactly the same parameters. There is some continuum background in the spectra of arcs, which varies from fibre to fibre and over time. This noise can interfere with line identification and the measurements of the line positions, especially for weak lines critical to our case.

To ensure consistent line identification, every arc spectrum is processed with a range of different thresholds for the background taken into the account in the line identification process. With different thresholds different numbers of lines are found. To find the lines Iraf uses our linelist and \texttt{center1d} algorithm to calculate the central wavelengths of arc lines. The threshold with which the expected number of lines are found is then adopted as a final solution. Detected lines are matched against our line list using our preliminary wavelength calibration (as explained in Section \ref{sec:nonauto}). A polynomial of the same shape as in the preliminary calibration is then fitted to the detected lines. Lines that are more than 2.5 sigma from the fit are rejected. Five such iterations are made and the wavelength calibration is then performed on the best matching lines. We limit the number of rejected lines to 15\% of the whole line list, however this limit is rarely reached. 

The fitted wavelength calibration is monitored periodically and if seen to be systematically deviating from the preliminary calibration,the spectrum is tagged, and is then skipped by the radial velocity and stellar atmospheric parameters estimator. This happens in $\sim$0.9\% of the cases for a variety of reasons: low throughput fibre, unpredicted background noise in arc spectra, problems with the arc lamp, and failed solutions due to an arc line remaining undetected in a critical part of the spectrum close to the edge of the range where the solution can not rely on any other line.

\subsection{Sky subtraction}
\label{sec:sky}
Out of 392 fibres available in each field, 25 are allocated as sky fibres. These fibres can be different in each field and are positioned as uniformly as possible over the field. Known stars are avoided, but galaxies or other objects are not, so it is possible for a fibre to be positioned at an inconvenient location. Thus not all 25 fibres can be trusted. Additionally, sometimes not all 25 fibres are allocated on a
field due to configuration limitations.

Our sky subtraction process first models the sky spectrum across the whole field and then subtracts it from all the spectra, as we now discuss.

\subsubsection{The sky spectrum}
On a moonless night the most prominent features in the sky spectrum are air glow lines, which are very strong in the IR and red arms, but there are only a few noticeable in the redward part of the green arm and none in the blue arm.

When the moon is up, the majority of the sky background in the sky spectrum comes from moonlight. Fields closer than 30$^\circ$ from the moon are avoided, so that moonlight contributes at most 120 counts per hour per pixel to the sky spectrum. 

\begin{figure}
\includegraphics[width=\columnwidth]{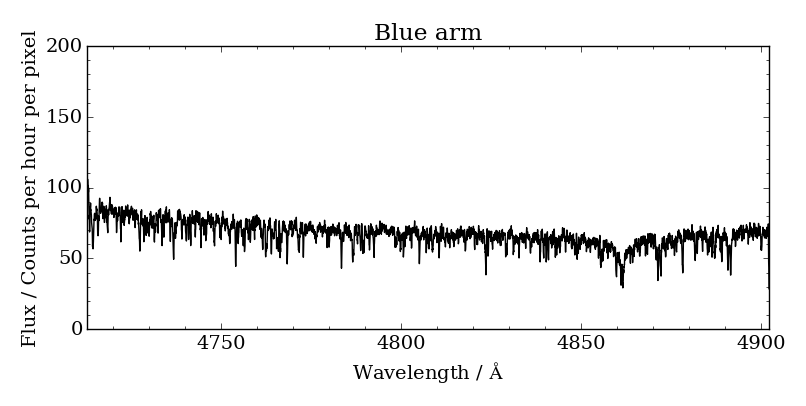}\\[-0.25cm]
\includegraphics[width=\columnwidth]{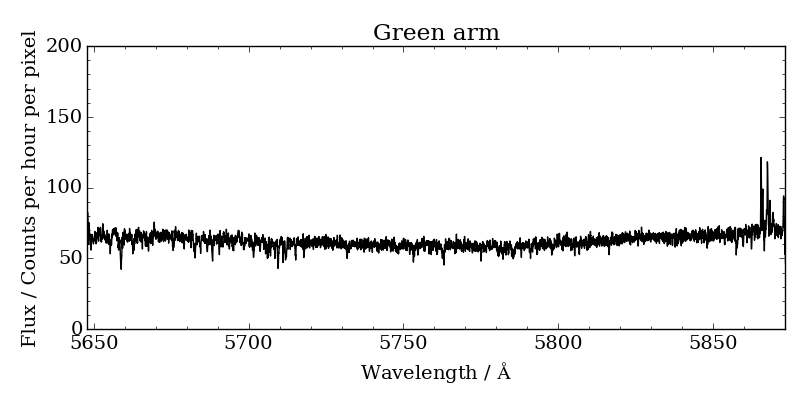}\\[-0.25cm]
\includegraphics[width=\columnwidth]{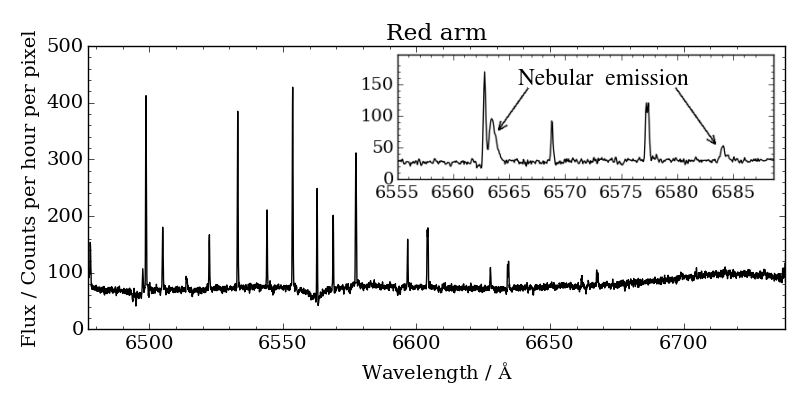}\\[-0.25cm]
\includegraphics[width=\columnwidth]{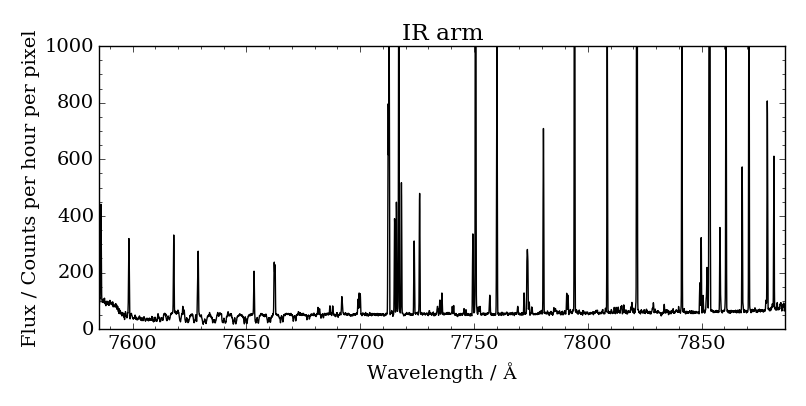}
\caption{Spectra of a moon-lit sky on June 6th 2015. A typical spectrum of the night sky is shown for all 4 arms. The main contribution to the sky spectrum in this case is the moon-light. Almost all of the emission lines come from the air glow. The inset is from a different night than the other spectra, when there was no moon above the horizon and in a field positioned near the Galactic plane, where nebular emission lines are present. Two marked lines are H$\alpha$ and [N II]. All spectra are an average of all the sky spectra in one field. There is residual flux of around 10-20 counts in all the spectra that appears smooth and for which we could not find a good explanation.}
\label{fig:sky_spec}
\end{figure}

In addition, we observed an unexplained flux of about 10 to 20 counts per hour per pixel (in blue and IR arm respectively). The spectrum of this contribution appears smooth. Despite our best efforts, we were unable to identify the source. It does not show any telluric absorption lines, so it must originate from somewhere near the telescope. This unexplained flux is best seen in the IR panel of Figure \ref{fig:sky_spec}, where saturated telluric absorption lines do not reach zero flux.

In a few fields close to the Galactic plane we observe emission lines coming from hot gas in the Milky Way itself. In these cases the sky subtraction cannot be trusted at wavelengths near these lines, as the nebulae of the Milky Way are much more structured than can be sampled by 25 sky fibres. Fortunately, the GALAH survey typically observes regions far away from the Galactic plane ($|b|>10^\circ$, see \citep{sarah2016}), and so such occurrences are very rare. Automated flagging of this condition has not been implemented in the pipeline, however the pipeline does provide a spectrum before the sky subtraction, so the user can decide whether or not the subtraction is still usable for their particular applications.

\subsubsection{Subtraction}
We need to be able to interpolate a sky spectrum over the whole 2$^\circ$ diameter field using 25 sky spectra (some of which may be contaminated by background sources). We must also account for the changing resolution delivered as a function of wavelength and position on the CCDs (see Section \ref{sec:ress}). 
The latter is largely solved by correcting the optical aberrations. Only a small variation of roughly 3\% remains (see Figures \ref{fig:res_map} and \ref{fig:res_map2}). 
This is small enough (compared to the intrinsic relative noise of the sky emission lines) that taking it into account is not necessary. 
Resolution changes more significantly with wavelength, but this change is the same for both sky fibres and object fibres so the subtracted sky spectra always have the correct resolution at every wavelength.

First we measure the fibre throughputs from a flat field. Excluding dead fibres, the throughputs of fibres can vary up to 50\% for different fibres \citep{jeffrey2016}. As noted above, the flat lamps do not illuminate fibres evenly; the illumination depends on the position and orientation of each fibre. However, a flat field is taken for every observed fibre configuration, so combining flat fields from different configurations is enough to average out the variations in the fibre illumination. Fibre throughputs are therefore measured from an averaged flat field and throughput is parametrised as a function of wavelength using 5th-order Chebyshev polynomials. The fibre with the highest throughput has its throughput set to one and all the spectra are normalized according to this scale with their corresponding polynomials. Absolute values of the throughputs are not relevant because we do not aim to produce flux calibrated spectra. 

Next, all 25 sky spectra are averaged with a rejection of values that are more than 3 sigma away from the average into a single spectrum (5 iterations are made). Spectra from ill-positioned fibres or any remaining cosmic rays, hot pixels, and vertical streaks are therefore removed from the averaged spectrum. This results in a high-SNR sky spectrum, so the subtraction will introduce less error.

\begin{figure}
\includegraphics[width=\columnwidth]{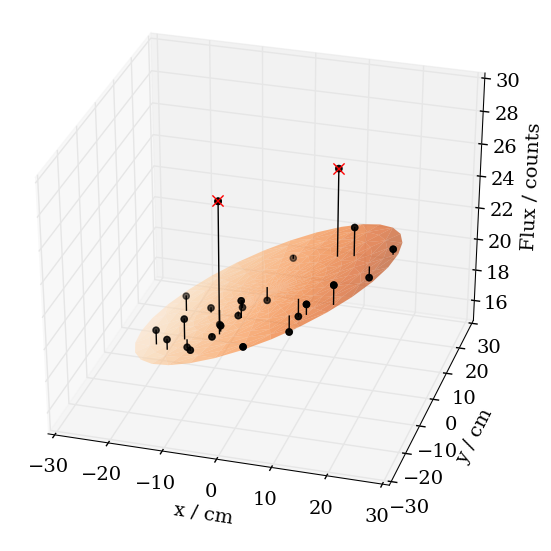}
\caption{Fit of a model sky plane to sky fibre fluxes.  Measured fluxes are represented by black points. After the fitting, two points were rejected and are crossed out in the image. The plane shows the size of the plate where fibres can be positioned. It is colour-coded by the mapped flux ranging from 16.5 counts/pixel to 23.8 counts/pixel. The $x$ and $y$ axes show position on the actual plate. This example shows red arm data.}
\label{fig:plane}
\end{figure}

We then map the sky flux across the whole field. The median flux in each sky fibre is measured and the positions of the fibres on the plate are read from the fibre configuration file. The sky flux is mapped by fitting a plane to the measured median sky fluxes. The plane is represented by three parameters: the tilt of the plane in two directions and the offset in the flux dimension. Fitting is done by minimizing the squared distances of points above or below the plane. Five iterations of the fit are performed, where points more than 3 sigma away from the plane are rejected in each step. This again removes outliers due to ill-positioned fibres. An example of such plane is shown in Figure \ref{fig:plane}.

We experimented with more complicated functions, like warped plane, where the tilt of the plane changes along one direction. This proved to be too complicated to fit automatically to 25 points while allowing the rejection of outlying points.

\begin{figure*}
\includegraphics[width=\textwidth]{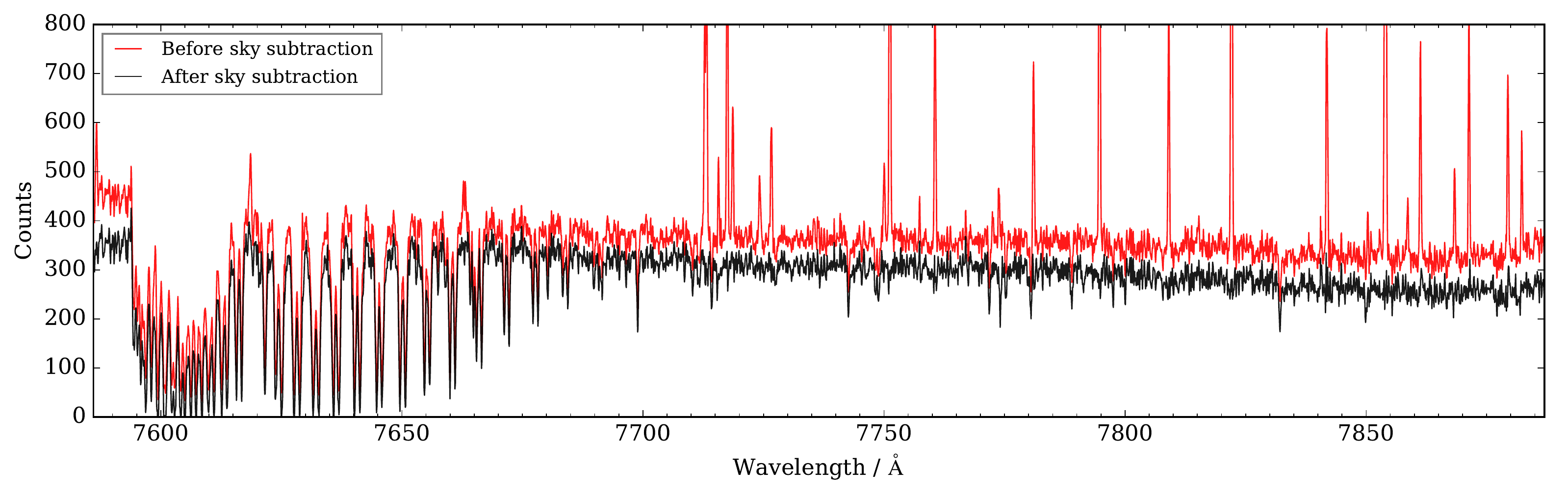}
\caption{IR arm spectra from September 9 2014 show the spectrum before the sky subtraction (red) and after the sky subtraction (black). The spectrum has a low SNR (SNR=25) so the sky emission lines are clearly visible. Note that also some continuum sky-light was subtracted.}
\label{fig:skysub}
\end{figure*}

The final step is then to take the position of each object fibre, calculate the model sky flux from the value of the model plane at that position, scale the high-SNR sky spectrum accordingly and subtract it from the spectrum for that fibre. Figure \ref{fig:skysub} shows an example of a sky subtracted spectrum in the IR arm where sky emission lines are strongest.

\subsection{Telluric absorption line removal}
Removal of telluric absorption lines usually requires additional observations of standard stars, which is time-consuming and not feasible for GALAH. It would also require us to model the absorption line profiles across the sky and with time to properly account for the effect of telluric absorption on GALAH science spectra. However, the GALAH wavelength ranges are impacted by telluric absorptions and do  have to be corrected. We correct GALAH science spectra by fitting the telluric lines directly in our observed spectra.

While telluric absorption lines are present in all four arms, they are only strong enough to be detectable and corrected in the red and IR arms. The blue end of the IR arm is in large part truncated by the A-band, a series of O$_2$ vibrational lines. Many of these lines are saturated and blended and are hard to correct for in a time efficient and automatic way. The IR arm wavelength range was selected so that the astrophysically interesting oxygen triplet lines lie in the centre of the range, which made the inclusion of a large part of the A-band. There are no features essential to the GALAH survey in the A-band. The only exception is the K I doublet, which lies in the less affected region with optically thin telluric lines. 

\begin{figure*}
\includegraphics[width=0.95\textwidth]{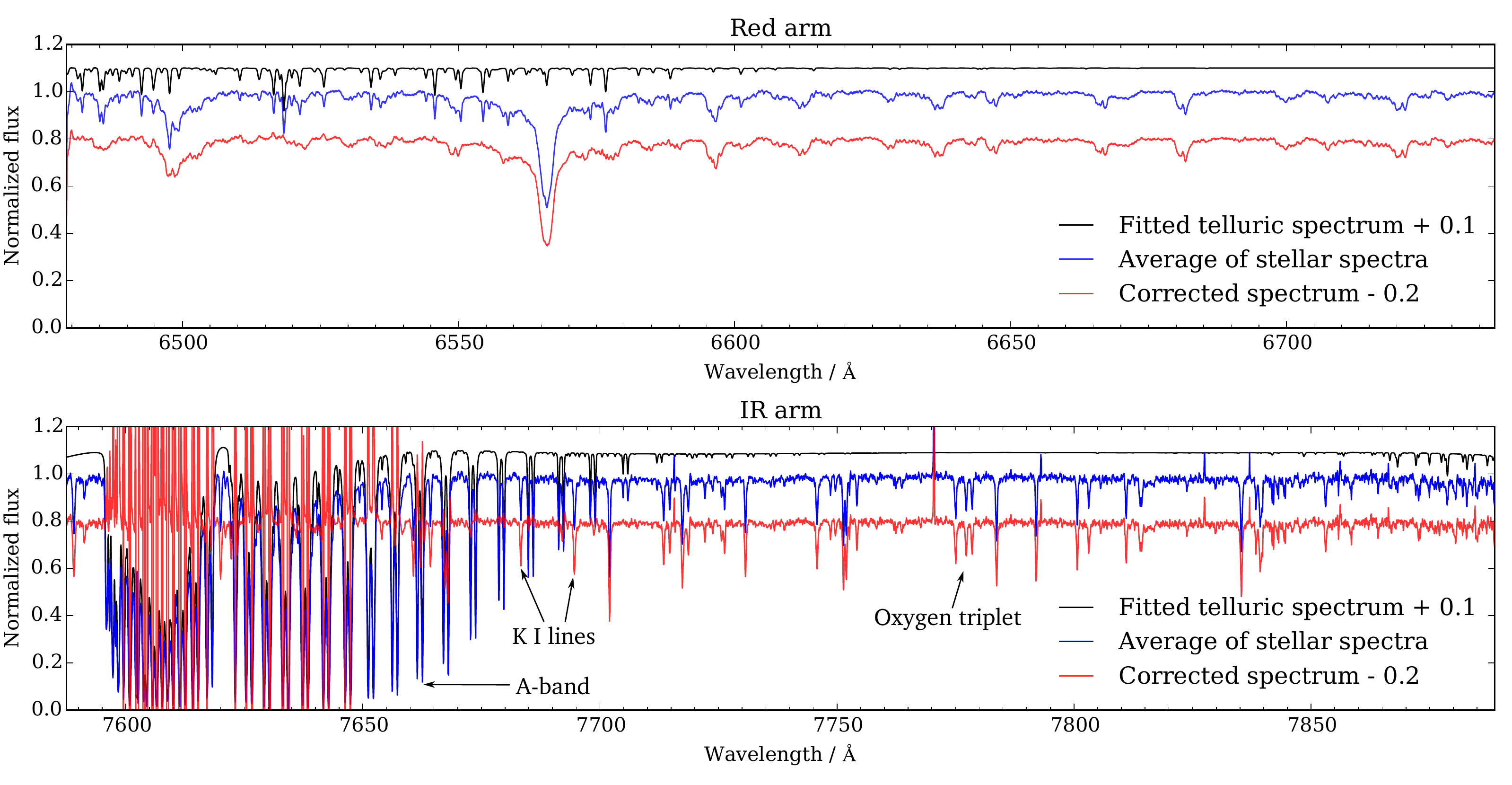}
\caption{Fitting and removal of telluric absorption lines. Plots show the fitted and corrected spectra in the red arm (top panel) and in the IR arm (bottom panel). The average spectrum of all objects in one image is plotted in blue. Black spectrum is the fitted telluric absorption spectrum. The corrected spectra are in red. The correction fails for lines in the A band that are optically thick but still works for the optically thin part of the A band. The plotted spectra are normalized for the purpose of this illustration. The process in the pipeline is done on non-normalized spectra. Features discussed in the text (Oxygen triplet, K~I lines and the A-band) are marked.}
\label{fig:tel}
\end{figure*}

Telluric absorption is fitted with ESO's Molecfit software \citep{kausch15, smette15}, release 1.1.1. Molecfit fits synthetic telluric absorption lines to spectra. To fit the telluric absorption spectrum to each spectrum takes around a minute on a top-end desktop computer, so it is infeasible to run Molecfit on every single spectrum. Instead, we first average all the spectra of objects in one image and fit the telluric absorption spectrum only once for an entire field. Our fields are observed at airmass low enough that no telluric line (with the exception of the A-band) will become optically thick in any part of the field. For each fibre we then calculate the difference between the average airmass and its actual airmass, and scale the fitted telluric spectrum accordingly. The individual spectra are then divided by scaled telluric spectra. An example of this process is shown in Figure \ref{fig:tel}.

\subsection{Barycentric velocity correction}
The wavelength calibration is adjusted for the barycentric velocity correction. The correction takes the effect of the Moon's motion and rotation of the Earth into account and calculates the projected velocity in the line-of-sight against the barycentre of the Solar system \citep{pyastro,pyastro-idl}. Already available routines proved to be fast and accurate in performing this step.

\subsection{Error spectra}
Our data start with an uncertainty determined by the photon-counting uncertainties in every pixel, and almost every step described above introduces further uncertainty and/or scales the spectrum away from a photon-counting distribution. We therefore propagate an initial photon-counting (i.e. Poisson) standard deviation for each pixel through the analyis to produce a so called error spectrum, or uncertainty spectrum that describes the standard deviation of each pixel in the final spectrum.

\section{Organization of reduced data}
\label{sec:org}
\subsection{Individual spectra}
Every reduced spectrum is named with a 16 digit number:
$$
\underbrace{140231}_{\mathclap{a}}\underbrace{0043}_{\mathclap{b}}
\underbrace{00}_{\mathclap{c}}\underbrace{125}_{\mathclap{d}}\underbrace{2}_{\mathclap{e}}.fits
$$
\begin{itemize}
\item[a)] Observation date in the format two digits for the year, two for the month and two for the day. This is the date when the night starts and does not change during the night.
\item[b)] Run number.
\item[c)] Combining algorithm. 00 in this case means that this is an uncombined individual spectrum (one out of three exposures).
\item[d)] three digit pivot number.
\item[e)] Arm number. 1 is blue, 2 is green, 3 is red, and 4 is the IR arm.
\end{itemize}

This name is unique and is used throughout, rather than star names or input catalogue numbers. Additionally, due to the exceptions like pilot programs and programs that do not use the GALAH selection function, but share the same observing routine, not all targets have input catalogue numbers.

There is one FITS file created per reduced spectrum. It contains four extensions: (i) the reduced spectrum itself, without any normalization; (ii) the error spectrum, which gives the relative uncertainty of the reduced spectrum. (iii) the spectrum without the sky subtraction and with no telluric absorptions fitted (apart from these two steps it is passed through the exact same steps in the reduction as the spectrum in the first extension); (iv) the relative error spectrum giving the uncertainty of the spectrum without the sky subtraction and with no telluric absorptions fitted. 

Until the final step of the reduction, each spectrum has a wavelength calibration saved as a polynomial giving a relation between pixels and wavelength. This is impractical in the next steps -- combining individual spectra and then estimating radial velocity and stellar atmospheric parameters, so the wavelength calibration is linearised by linearly interpolating the values between pixels. Median dispersion is used for the linearised wavelength calibration, so the number of pixels per spectrum remains roughly the same after this process. The wavelength calibration is the same for all four extensions. It is linearised and in a standard FITS format given by NAXIS1, CRPIX1, CRVAL1, CRDELT1, and CUNIT1 for the number of pixels, the reference pixel, wavelength in the centre of the reference pixel, dispersion per pixel, and units, respectively.

\subsubsection{Combined spectra}
While usually three consecutive 20-minute exposures are made for each field, this is not always the case as there can be more, or less, depending on weather conditions (as decided by the observer at the telescope). A flat field and arc frame are also made with the same fibre setup as the science exposures for each field. This continuous block of exposures is assigned a number called the cob\_id (continuous observational block identifier). A combination of the science exposures from this block makes one observation used for further scientific analysis. Individual uncombined spectra are retained, but usually not used for science purposes. The cob\_id is simply the combination of the 6-digit date marker and the run number of the first exposure made in this block, regardless of its type (flat, arc or science exposure). If the same field is observed in two parts, with other fields being observed in between, two cob\_ids will be assigned and they will be treated as two separate observations. The cob\_id also changes if the same objects are observed with different plates. Even if the objects are the same, the field configuration is different, so such observations are treated separately.

A new name is assigned to combined spectra:
$$
\underbrace{1402310042}_{\mathclap{a}}\underbrace{01}_{\mathclap{c}}\underbrace{125}_{\mathclap{d}}\underbrace{2}_{\mathclap{e}}.fits
$$
Instead of the date and run marker, we here use the cob\_id (marked with \textit{a}). Note the difference in the \textit{c} section of the name. Number 01 indicates that this is a combined spectrum and the method used for combining all the science exposures was the method number 1. This also implies that the cob\_id is used in the first part of the name.

The structure of the FITS file is the same as for the individual spectra, with the addition of a list of spectra that were combined in each header.

Currently we use two combining methods, which are reflected in the file name:
\begin{enumerate}
\item[1:] All the science exposures of the same object with the same cob\_id are summed together. Error spectra are combined appropriately, so they still give relative uncertainties.
\item[2:] All the science exposures with the same cob\_id are combined as in the case 1 above. Additionally, all the exposures of the same object made at earlier times are also added, even though they have different cob\_ids. Error spectra are combined appropriately, so they still give relative uncertainties. We emphasize here that this combines an exposure of an object
with all previous exposures. The cob\_id of this output
spectrum is the cob\_id of the last combined exposure. For
example, if an object is observed on 3 days, there are two
resulting spectra with combining methods set to 2. The first
combines the exposures from the first and second day and
carries the cob\_id of the second day. The second output
spectra combines the exposures from all three days and has
the cob\_id of the third day. To get the version of the spectrum
that combines all previous exposures, select that with the
latest cob\_id.
\end{enumerate}

Sometimes another number called sobject\_id is used. It is the same as the name of the spectrum described above, but drops the last digit (denoting the arm), so it stands for the object as a whole and not for individual spectra for each arm. An sobject\_id for the above example will therefore be:
 $$
\underbrace{1402310042}_{\mathclap{a}}\underbrace{01}_{\mathclap{c}}\underbrace{125}_{\mathclap{d}}
$$
where \textit{a}, \textit{c}, and \textit{d} mark the same parts of the name as before.

\section{Estimation of radial velocities and stellar atmospheric parameters}
\label{sec:rv}
After the reduction of spectra we run a code that provides first estimates of radial velocity and three basic atmospheric parameters for each star. The spectra are delivered to the collaboration only after this step is performed, so we present this software code together with the reduction pipeline. Radial velocities and stellar parameters are calculated using the blue, green and red arms only. The fourth arm is omitted as it contains significant amount of telluric contamination and comparatively less parameter-sensitive information than the other three arms.

Radial velocities are calculated by cross correlating the observed spectrum with a set of 15 AMBRE model spectra \citep{laverny12}. Each arm is individually cross correlated with the grid, one model at a time. The model set spans 4000--7500~K in temperature, with 250~K intervals. It has fixed $\log g$ and metallicity at 4.5~dex and 0~dex, respectively. 

Normalization is done by fitting a univariate spline function over the entire observed spectrum. Areas of prominent spectral features (such as the H$\alpha$ and H$\beta$ lines) are omitted from the fitting. Both the model and the observed spectra are then moved to zero level by subtracting their means and are multiplied by a window function to smooth out their edges. This is done to reduce the noise in the cross correlation peak. The normalized spectrum is then cross correlated with the 15 models to determine the radial velocity from this particular arm. Cross corelation function is calculated in steps of one pixel, so the exact radial velocity is calculated by fitting the highest peak in the cross corelation function. The cross correlation peak is fitted with a 2nd degree polynomial and we adopt the maximum of the fit as the cross correlation coefficient.  Only model spectra which have cross correlation coefficient greater than 0.3 (the coefficient ranges from 0 to 1, with 1 being the perfect correlation) are included in the radial velocity determination. 

\begin{figure} 
\centering
\includegraphics[width=\columnwidth]{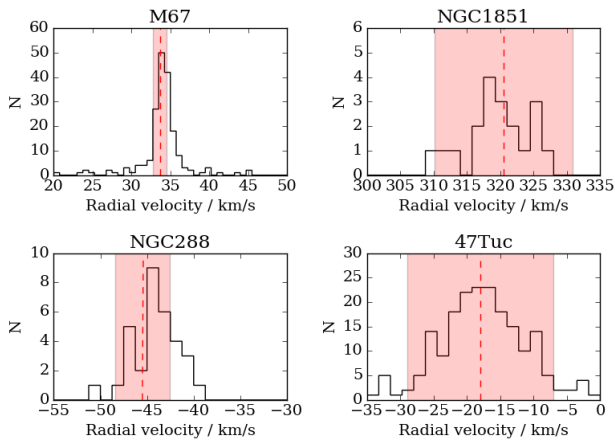}
\caption{Radial velocity distributions of four well studied clusters. The red dashed lines are literature values, the red boxes indicate their associated dispersions. For each cluster we first give the mean radial velocity and dispersion from the literature and then our own values. Top left panel: M67, 195 stars. Literature values: 33.64 and 0.854~$\mathrm{km\ s^{-1}}$ \citep{geller15}, this work: 33.71 and 8.02~$\mathrm{km\ s^{-1}}$. Top right panel: NGC1851, 19 stars. Literature values: 320.5 and 10.4~$\mathrm{km\ s^{-1}}$ \citep{harris96}, this work: 319.52 and 4.75~$\mathrm{km\ s^{-1}}$ . Bottom left panel: NGC288, 32 stars. Literature values: $-45.4$ and 2.9~$\mathrm{km\ s^{-1}}$ \citep{harris96}, this work: $-44.22$ and 2.28~$\mathrm{km\ s^{-1}}$. Bottom right panel: 47~Tuc, 196 stars. Literature values: $-18$ and 11~$\mathrm{km\ s^{-1}}$ \citep{harris96}, this work:$-17.42$ and 6.61~$\mathrm{km\ s^{-1}}$. It is evident that our radial velocities match nicely the literature radial velocities for these four clusters. We attribute the discrepancies in the radial velocity dispersions to contamination by field stars (in M67), small number of observed stars (in NGC1851), and cluster centres not being included in GALAH observations (NGC1851 and 47Tuc).}
\label{fig:rv}
\end{figure}

The radial velocity for each arm is a combination of the acceptable model spectra, weighted by their cross correlation coefficient. The radial velocities from all three arms are averaged to give the final radial velocity, and the standard deviation between the arms is adopted as the uncertainty. The uncertainty is not derived from the peak width of the cross correlation function, because of large dispersion and cannot be calculated accurately from a sample of only 15 cross correlation functions. If one arm has a radial velocity lying more than two times further than the difference between the other two, it is excluded from the final radial velocity. Figure~\ref{fig:rv} shows the comparison between radial velocities from this work and from literature for four well studied clusters. With the exception of M67, our radial velocities for each cluster agree with literature values within their cited uncertainties. For M67, the peak of our velocities agree well with the literature value, however, the observed dispersion is large compared to literature values because it includes possible non-members, an unknown fraction of spectroscopic binaries, and has not been corrected for the individual velocity errors, which for this cluster, are comparable with the intrinsic cluster velocity dispersion.

\begin{figure}
\includegraphics[width=\columnwidth]{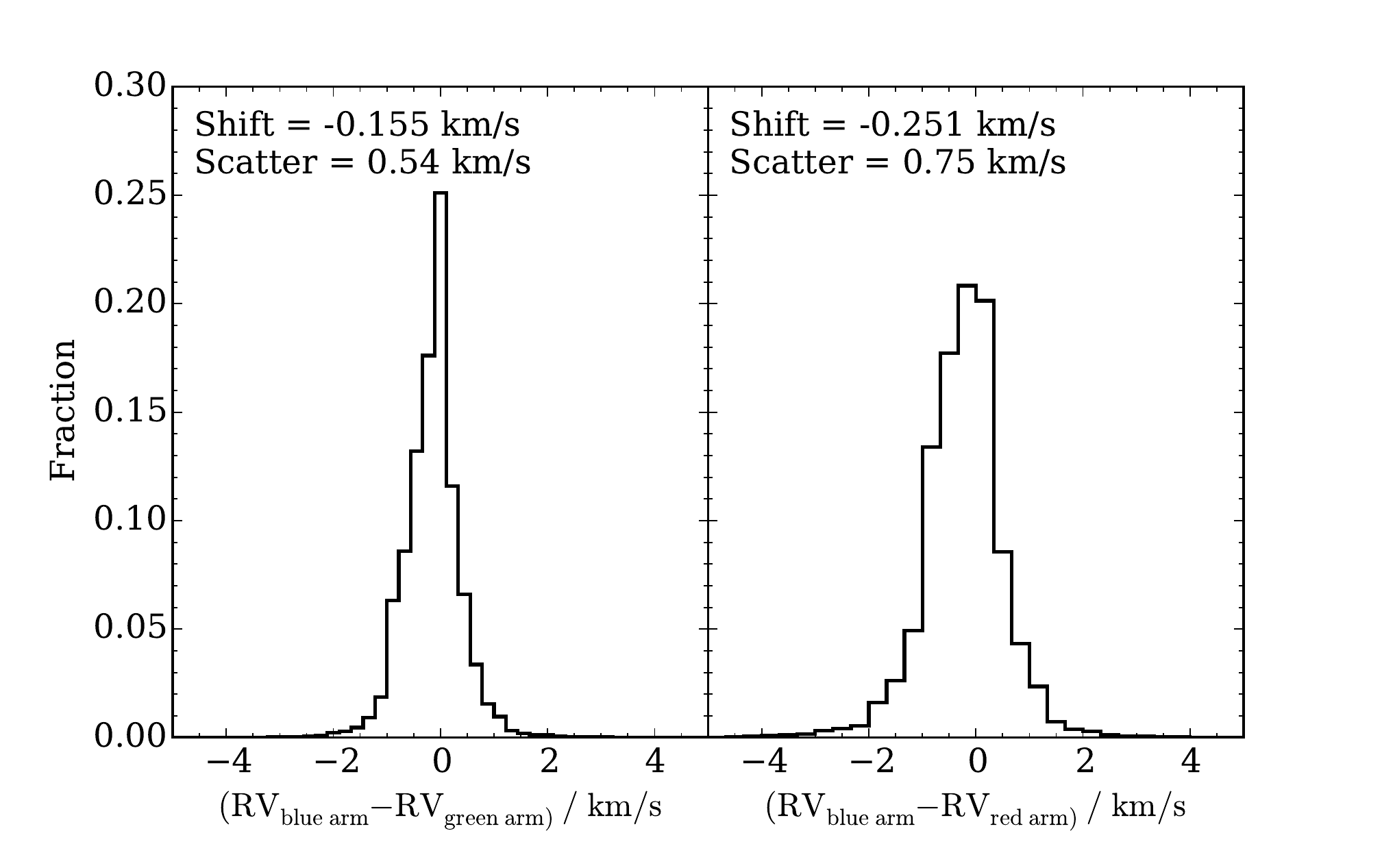}
\caption{Difference between the blue arm and green arm radial velocity (left panel) and the difference between the blue and red arm radial velocity (right panel). Radial velocity is not measured from the IR arm. Shift from zero difference and scatter of the distribution are given in each panel. Radial velocities of $\sim124000$ stars were used to produce this plot.}
\label{fig:rv_dif}
\end{figure}

Figure~\ref{fig:rv_dif} shows the statistics for the differences between the radial velocities measured in each arm. There is a systematic shift between each arm. It does not correspond to the shift we see in Figure~\ref{fig:linelists2}, concluding that the shifts shown in Figure~\ref{fig:linelists2} are a consequence of imprecise literature values for a large number of lines. We attribute the non-zero radial velocity differences in Figure~\ref{fig:rv_dif} to systematic errors in the radial velocity calculation scheme and the initial ThXe linelist we composed, which issues are propagated to every radial velocity calculated. However, the shift is almost two orders of magnitude smaller than the resolution of the spectrograph, meaning that our wavelength calibration method works. 

Stellar parameters (T$_{\mathrm{eff}}$,$\log g$ and [Fe/H]) are derived by globally fitting of the observed spectra with a grid of 16783 AMBRE spectra \citep{laverny12} , which are convolved down to the average resolution of each CCD. Tables~\ref{tab:wltable} and~\ref{tab:ambretable} contain the wavelength and parameter ranges of this grid. The aim of this procedure is to provide initial stellar parameters, which are used as the starting model to help speed up the stellar parameters and abundance pipeline. This global fitting method is both robust and efficient, taking under a second per star to converge. 

\begin{figure}
\includegraphics[width=\columnwidth]{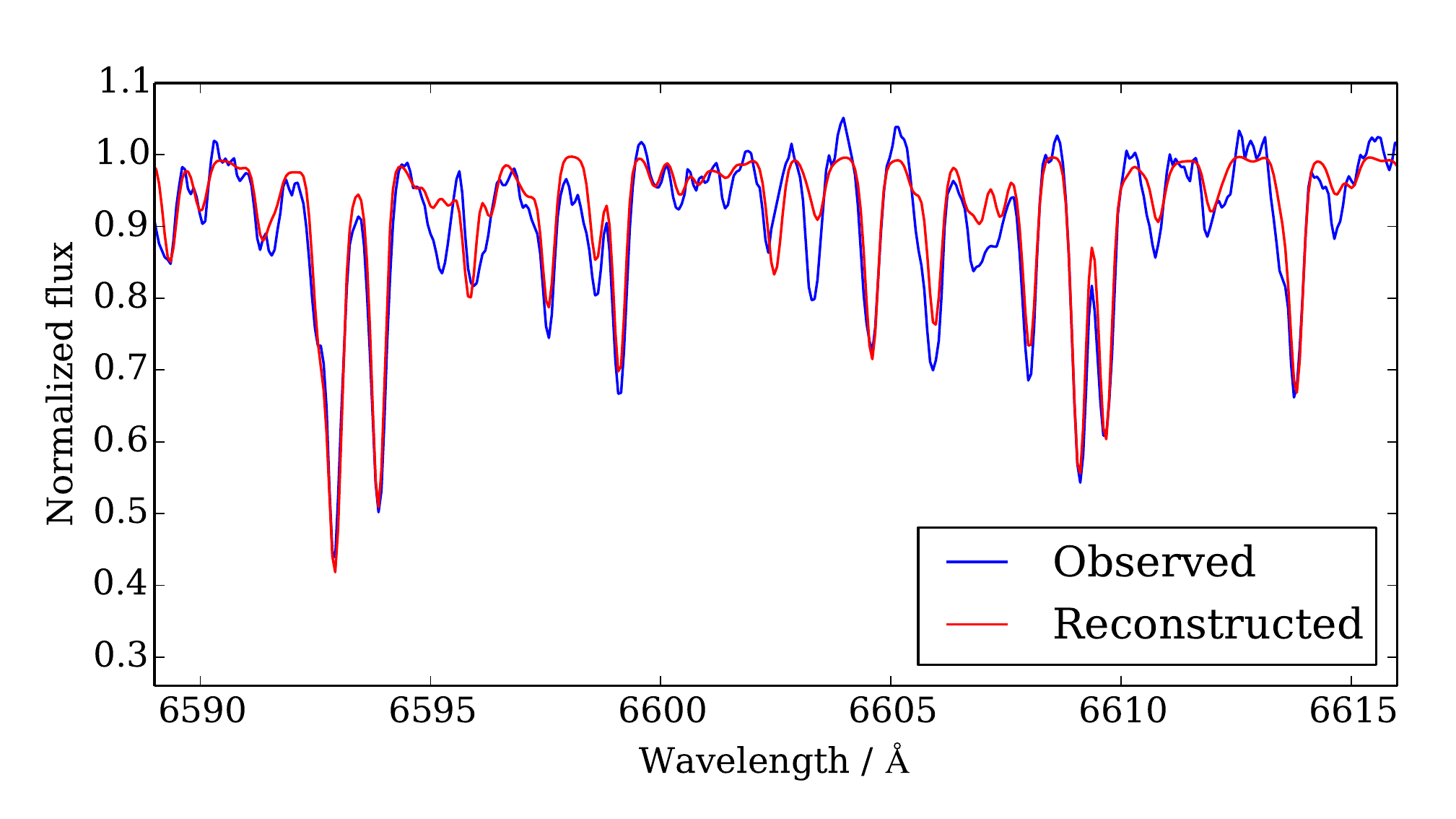}
\caption{The comparison between the reconstructed (red) and the observed (blue) spectra for a star. Most spectral features are well reproduced, with only a few lines (e.g., line at 6603.5 $\rm \AA$) with erroneous strengths. Such deviations are caused by inacurate linelists used to produce the synthetic spectra.}
\label{fig:recon}
\end{figure}

The radial velocity corrected spectrum is normalized by fitting 3rd (for blue and green arms) and 4th (for red and IR arm) order polynomials to predetermined continuum regions. These regions were determined by examining high resolution spectra of the Sun, Arcturus and $\mu$~Leo. Normalized spectra are interpolated on to the common wavelength grid as the model. We select the ten closest-model spectra in Euclidean space to the observed spectrum (by computing the $\rm L^2$ norm between the observed spectrum and each model spectrum) and reconstruct the original spectrum using a linear combination of these ten. The stellar atmospheric parameters are the same linear combinaction of the parameters of the ten closes-model spectra. Ten spectra proved to be a safe number to use considering the size of the synthetic grid and quality of our data. Figure~\ref{fig:recon} shows an example of the reconstructed and the observed spectra.

\begin{figure} 
\centering
\includegraphics[width=\columnwidth]{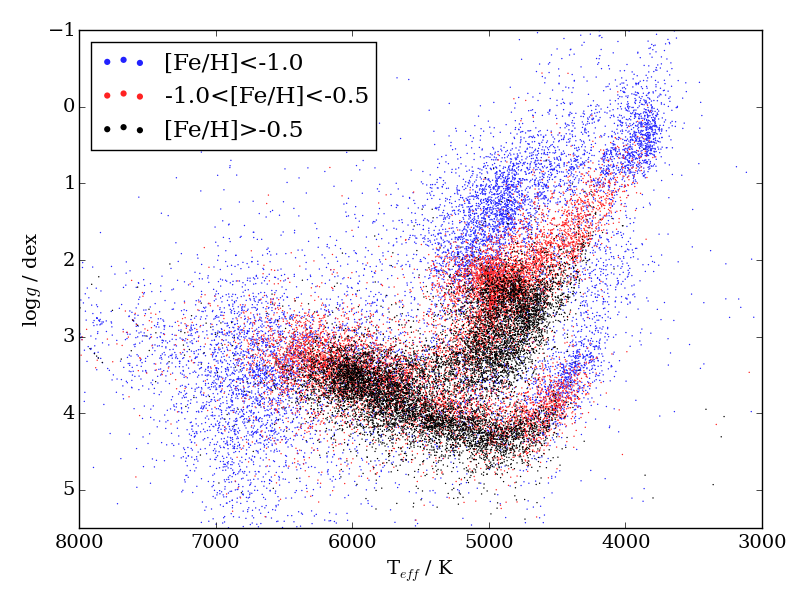}
\caption{The spectrometric Kiel diagram of 10000 randomly selected GALAH stars with metallicity binned into three bins.} 
\label{fig:hr}
\end{figure}

Figure~\ref{fig:hr} shows the Kiel diagram of 10000  GALAH stars with parameters derived from this method. We note the erroneous upturn at the lower part of the main sequence, a common issue encountered by global fitting algorithms caused by the breakdown of the one dimensional local thermal equilibrium models in cool dwarfs \citep[e.g.][]{worley16}. 

\begin{table}
\begin{center}
\begin{tabular}{ c | c | c |c}
\hline
  	CCD & colour & wavelength / \AA & sampling interval / \AA\\ \hline
  1 & blue & 4717 -- 4887 & 0.0453\\
  2 & green & 5651 -- 5864 & 0.0546\\
  3 & red & 6483 -- 6729 &0.0623\\
\hline
\end{tabular}
\end{center}
\caption{Wavelength coverage and sampling for the model grid.}
\label{tab:wltable}
\end{table}

\begin{table}
\begin{center}
\begin{tabular}{  c | c |  c}
\hline
parameter & range & increments\\ \hline 
  \multirow{2}{*}{$T_{\rm eff}$ / K} & \multirow{2}{*}{$2500$ -- $8000$} & $200$ ($T_{\rm eff}$ $\rm < 4000$)\\ 
  & & $250$ ($T_{\rm eff}$ $\rm > 4000$)\\[0.2cm]
  $\log$g / $\rm \log\left(cm s^{-2}\right)$ & $-0.5$ -- $5.5$ & $0.5$ \\[0.2cm]
  \multirow{3}{*}{$\rm [Fe/ H]$ / dex} & \multirow{3}{*}{$-5$ -- 1} & $1$ ($\rm [Fe/ H] <-3$)\\
  & & $0.5$ ($-3<\rm{[Fe/ H]} <-1$)\\
  & & $0.25$ ($-1<\rm{[Fe/ H]}$) \\ 
\hline
\end{tabular}
\end{center}
\caption{Parameter summary of the model grid.}
\label{tab:ambretable}
\end{table}

\section{Resolution maps}
\label{sec:ress}

Resolution of reduced spectra is not self-evident, despite the information provided by the instrument designers \citep{sheinis15}. The star light passes through many optical elements and computational processes, so it is not trivial to calculate the resolution of final reduced spectra. It has to be measured from actual observed spectra. It is, however, critical that the real resolution is known by analysis methods. We therefore measured the resolution by two methods: from arc spectra and from spectra produced by an optical comb \citep{steinmetz08}.

\subsection{Resolution maps from arc lines}
Arc spectra consist of unresolved emission lines, conveniently distributed over most of the observed wavelengths. Arc lines are strong enough to have their profile measured in individual spectra. The alternative is to use sky emission lines, also unresolved but weaker than arc lines and found only in parts of the red and IR arm.

\begin{figure}
\includegraphics[width=\columnwidth]{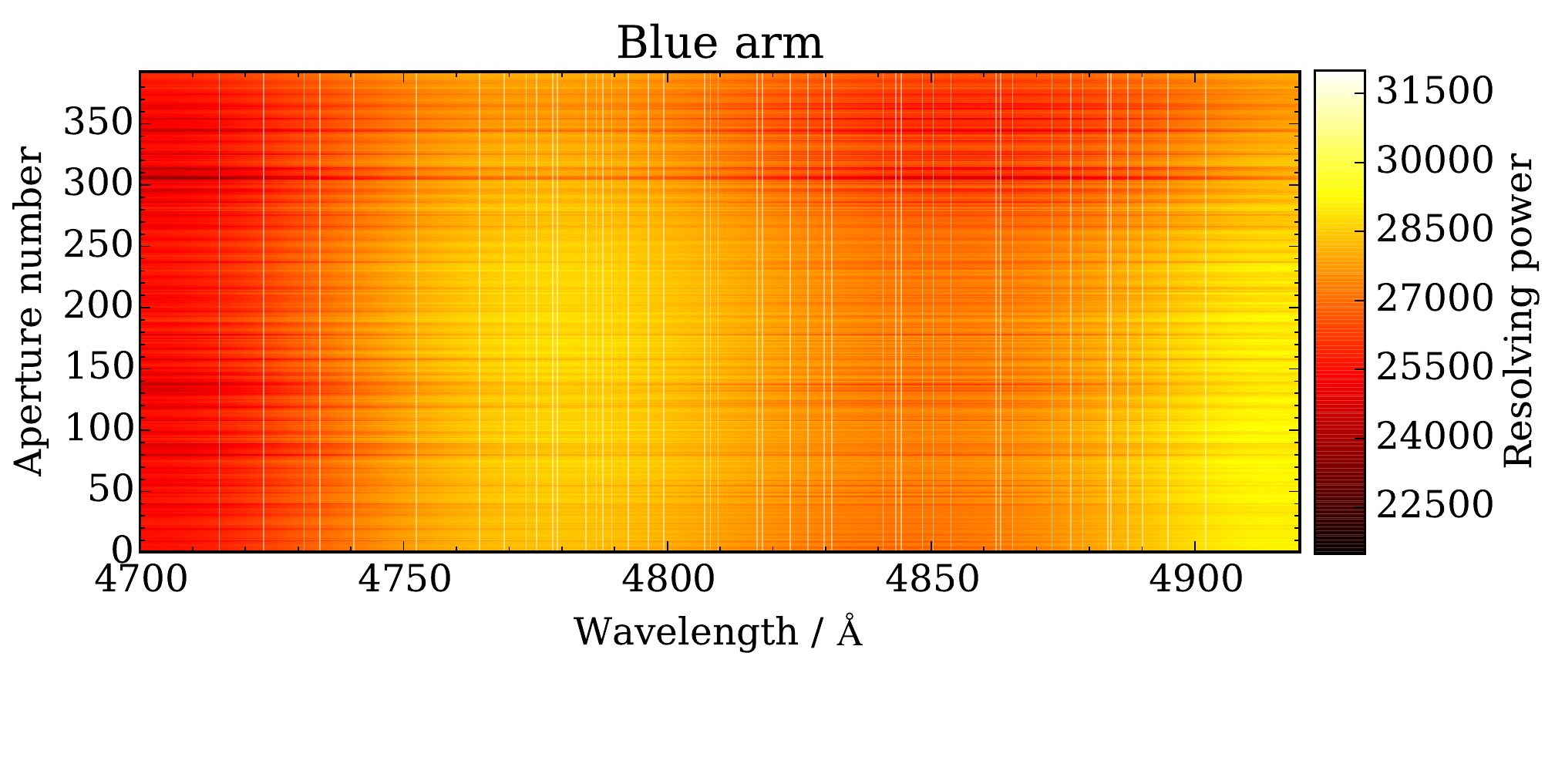}\\[-0.5cm]
\includegraphics[width=\columnwidth]{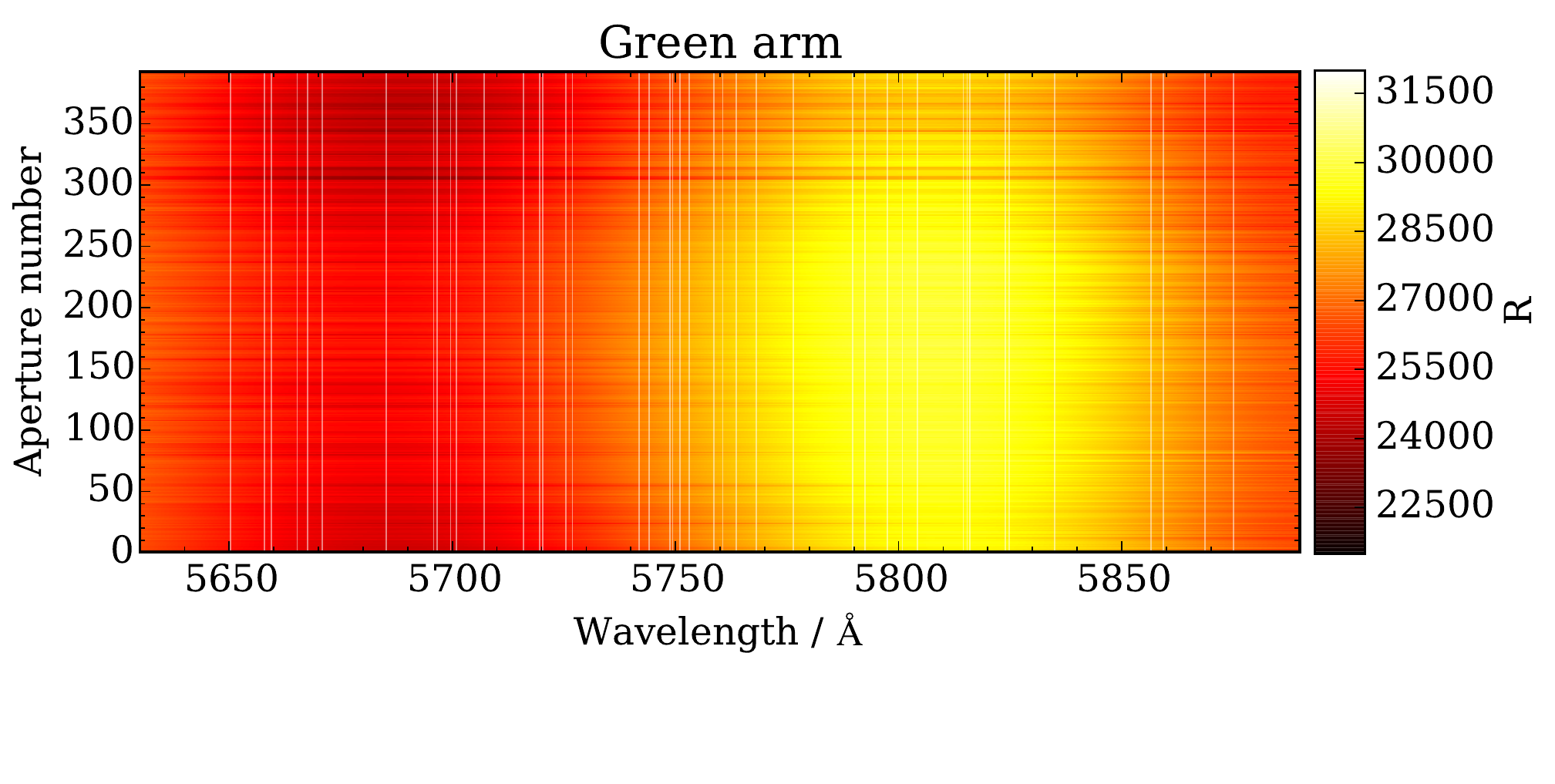}
\caption{Resolution maps for the blue arm (top) and the green arm (bottom). The maps show the colour coded resolving power as it depends on the wavelength and aperture number. White vertical lines show arc lines that were used to produce these maps. Map is composed of 6th degree polynomials that represent the resolving power for each aperture. Polynomials were fitted to measured arc lines widths for each aperture independently.}
\label{fig:res_map}
\end{figure}

\begin{figure}
\includegraphics[width=\columnwidth]{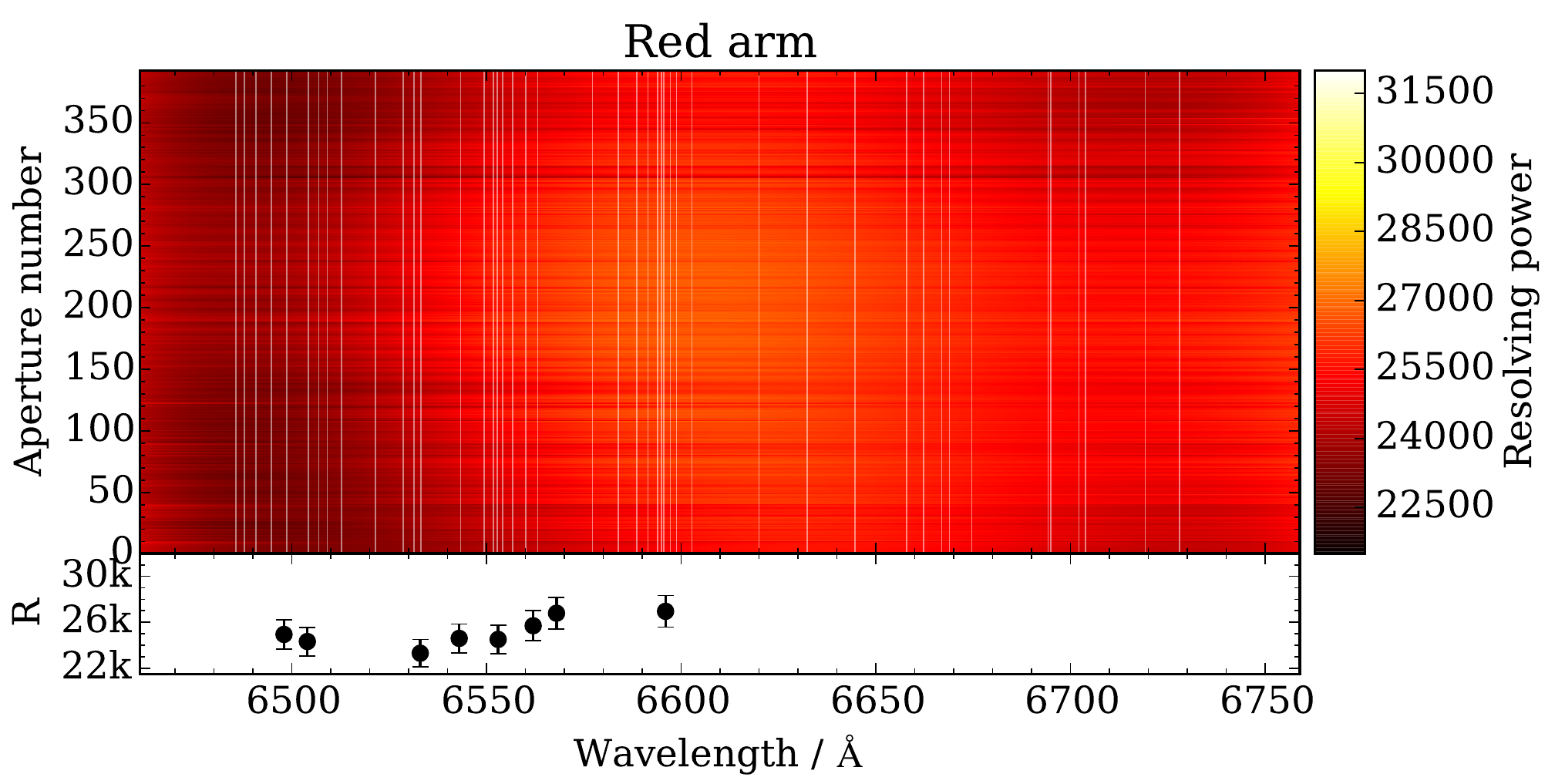}\\
\includegraphics[width=\columnwidth]{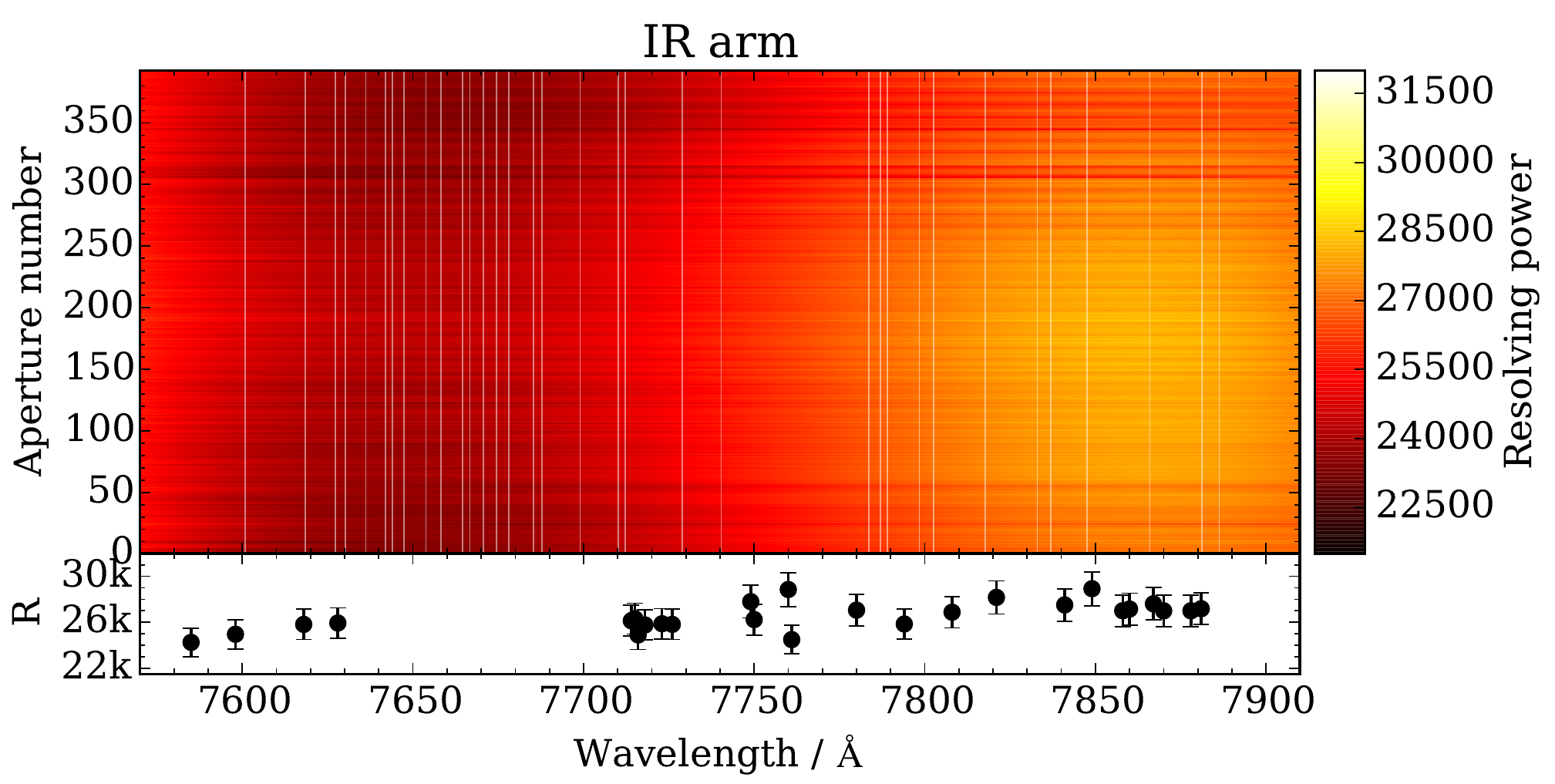}
\caption{Resolution maps for the red arm (top) and the IR arm (bottom). The maps show the colour coded resolving power as it depends on the wavelength and aperture number. White vertical lines show arc lines that were used to produce these maps. Map is composed of 6th degree polynomials that represent the resolving power for each aperture. Polynomials were fitted to measured arc lines widths for each aperture independently. The smaller panels below the resolution map show how the resolving power changes with wavelength when measured from sky emission lines. Each dot represents the resolving power measured from one line and the errorbars show the scatter accumulated over many images. There are no detectable sky emission lines in the blue and green arm. The resolution in the case of sky emission lines was measured from sky spectra combined from all the sky fibres in the image, so no dependence on aperture number is given.}
\label{fig:res_map2}
\end{figure}

We measured the full width at half maximum (FWHM) for each arc line in all the allocated fibres for over 200 images randomly selected from the whole survey campaign. The resolution maps show the resolving power:
\begin{equation}
R=\frac{\lambda}{FWHM},
\end{equation}
which was calculated for each measured line. A 6th degree polynomial was fitted to all the measurements in each aperture. Because some arc lines are blends of two unresolved lines, a rejection algorithm was used to fit the polynomial only to measurements of single lines. A polynomial was fitted independently for each aperture and the result is plotted in Figures \ref{fig:res_map} and \ref{fig:res_map2}. 

Due to concerns that the emerged pattern is distorted by blends of unresolved arc lines, we tried to verify the measured resolution from measurements of profiles of sky emission lines. Because sky emission lines are weak in individual spectra, we used combined sky spectra, like the one in Figure \ref{fig:sky_spec}. The result is shown in Figure \ref{fig:res_map2} under the red and IR maps. There are no usable sky emission lines in the green and blue arms and only a few in the red arm. We observe a correlation in the resolution measured in both ways, so we believe that the measured resolution profiles are the consequence of the instrument and not of inconveniently blended arc lines. We test this further with the help of a photonic comb in the following section.

The measured resolution is in a rough agreement with the resolving power of 28000, for which the instrument was designed. The exception is the red arm, where the lower average resolution is due to out-of-focus CCD, a mechanical error that will be fixed during the planned maintenance of the HERMES spectrograph in late 2016.

\subsection{Resolution maps from the photonic comb}

In March and April 2016 we were given 4 days at the AAT to try coupling a photonic comb to the 2dF for the purpose of measuring resolution, check our method of wavelength calibration and test the capability of a photonic comb to serve as a calibration for observations in the future. We were unable to take data of sufficient quality to challenge the wavelength calibration method presented in Section \ref{sec:wav}, but the data was suitable for the resolution measurements. 

The photonic comb spectrum is generated by coupling white light into a single-mode fibre based broadband Fabry-Perot etalon manufactured to have a finesse (ratio of the etalon's transmission peak separation to the width of the peaks) of 40 and a 453~$\mu$m cavity.
This produces a peak separation of 4~{\AA} at 7800~{\AA} and 2.5~{\AA} at 5700~{\AA} both with a width of approximately 1.1~{\AA}. The etalon is enclosed with a small thermoelectric cooler that is used to stabilise the transmission spectrum. The small size of the cavity allows for relatively quick changes (multiple degree changes can settle in less than a second). To stabilise the photonic comb spectrum we use a Doppler-free saturation absorption spectroscopy setup \citep{1996AmJPh..64.1432P} that allows simultaneous measurement of the rubidium \emph{D$_2$} hyperfine transition lines and a transmission line of the photonic comb using a tunable laser. This rubidium transition is known to be extremely stable and is an excellent absolute wavelength reference. The photonic comb spectrum is transmitted to the spectrograph using a series of single-mode fibre switches to swap the input and output of the etalon between a white light source or the tunable laser and to feed either the spectrograph or Rb locking setup, respectively. As a precaution the setup was limited to monitoring the etalon in-between exposures to avoid coupling the tunable laser into the spectrograph. These measurements indicated stability on the order a few m/s over each exposure. 

The photonic comb produces Lorentzian profile lines with a FWHM of 0.15~{\AA} in the IR arm, which is less than the resolution. However, the wide Lorentzian wings are resolved. Even more, they are detectable in the whole region between two peaks in a reasonably exposed spectra with more than a few thousand counts in the peaks. When measuring the resolution the spectra should be deconvolved with the light source spectral profile. Resolving the Lorentzian profiles allows us to skip this step and fit Voigt profiles (a convolution of a Gaussian and Lorentzian profile) instead. The Voigt profile can be conveniently written as:
\begin{equation}
\begin{aligned}
V(x;A,x_0,\gamma,\sigma)=A\frac{\mathrm{Re}\left( w\left(z\right)\right)}{\sigma\sqrt{2\pi}},\\
w(z)=\exp\left( -z^2\right)\mathrm{erfc}(-iz),\\
z=\frac{(x-x_0)+i\gamma}{\sigma\sqrt{2}},
\end{aligned}
\end{equation}
where $A$ is the amplitude, $x_0$ is the centre of the profile, $\gamma$ is the parameter of the Lorentzian part, and $\sigma$ is the parameter of the Gaussian part of the Voigt profile. The resolving power can then simply be calculated by adopting $\sigma$:
\begin{equation}
R=\frac{\lambda}{2.3548\ \sigma}
\end{equation}

\begin{figure}
\includegraphics[width=\columnwidth]{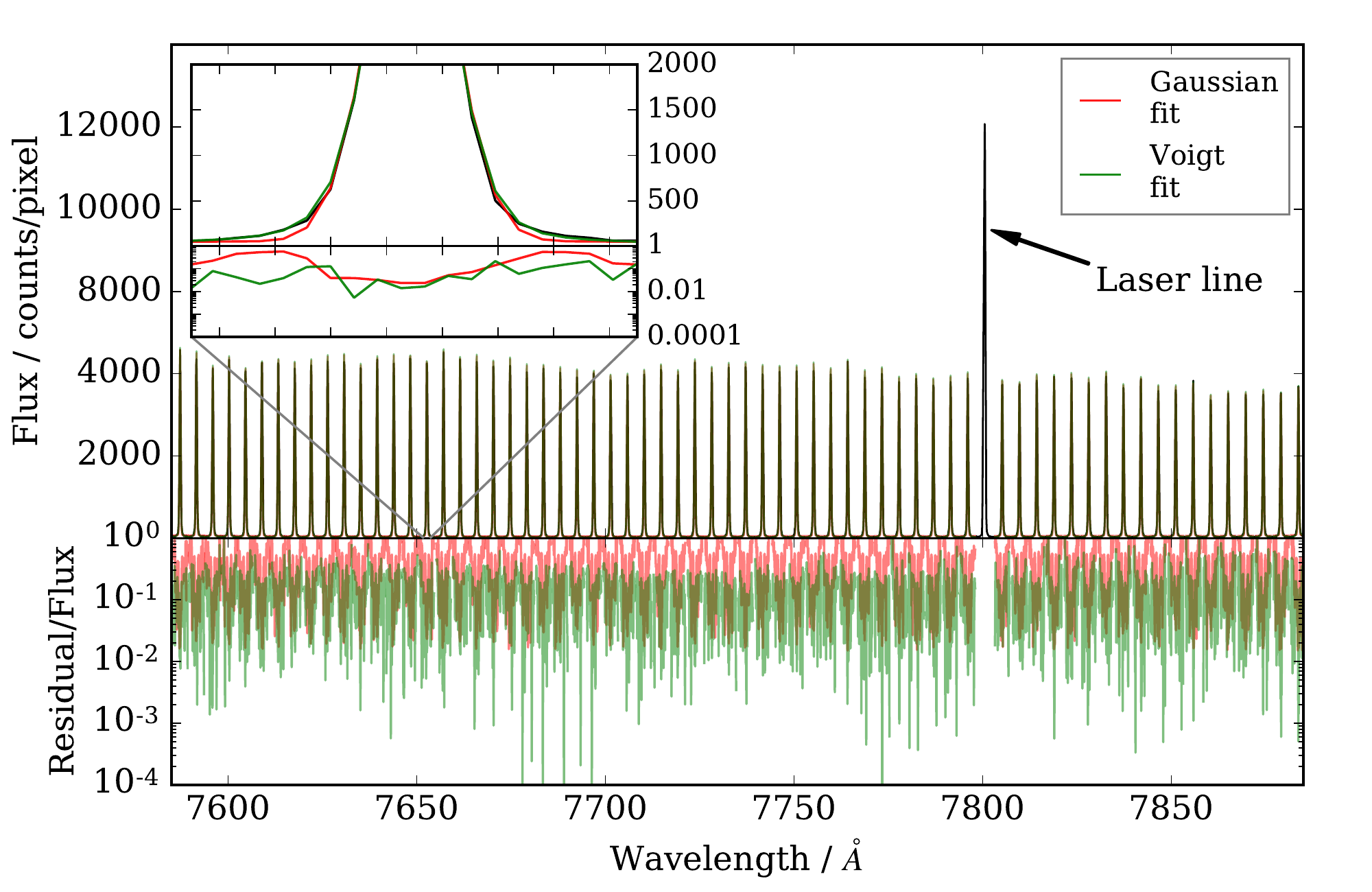}
\caption{Spectrum produced by the photonic comb. The marked line is the Rb laser line. The laser does not shine directly into the fibre, but due to its strength some light scatter of the dome and into the fibres. The inset shows one typical line in detail. Red and green lines show fitted Gaussian and Voigt profiles. Only the IR arm spectrum is shown here. Green and Red arms spectra look similar, but have somewhat less signal and no laser line of course. Blue arm spectra are useless for our application because the photonic comb produces too weak signal at those wavelengths.}
\label{fig:comb1}
\end{figure}

\begin{figure}
\includegraphics[width=\columnwidth]{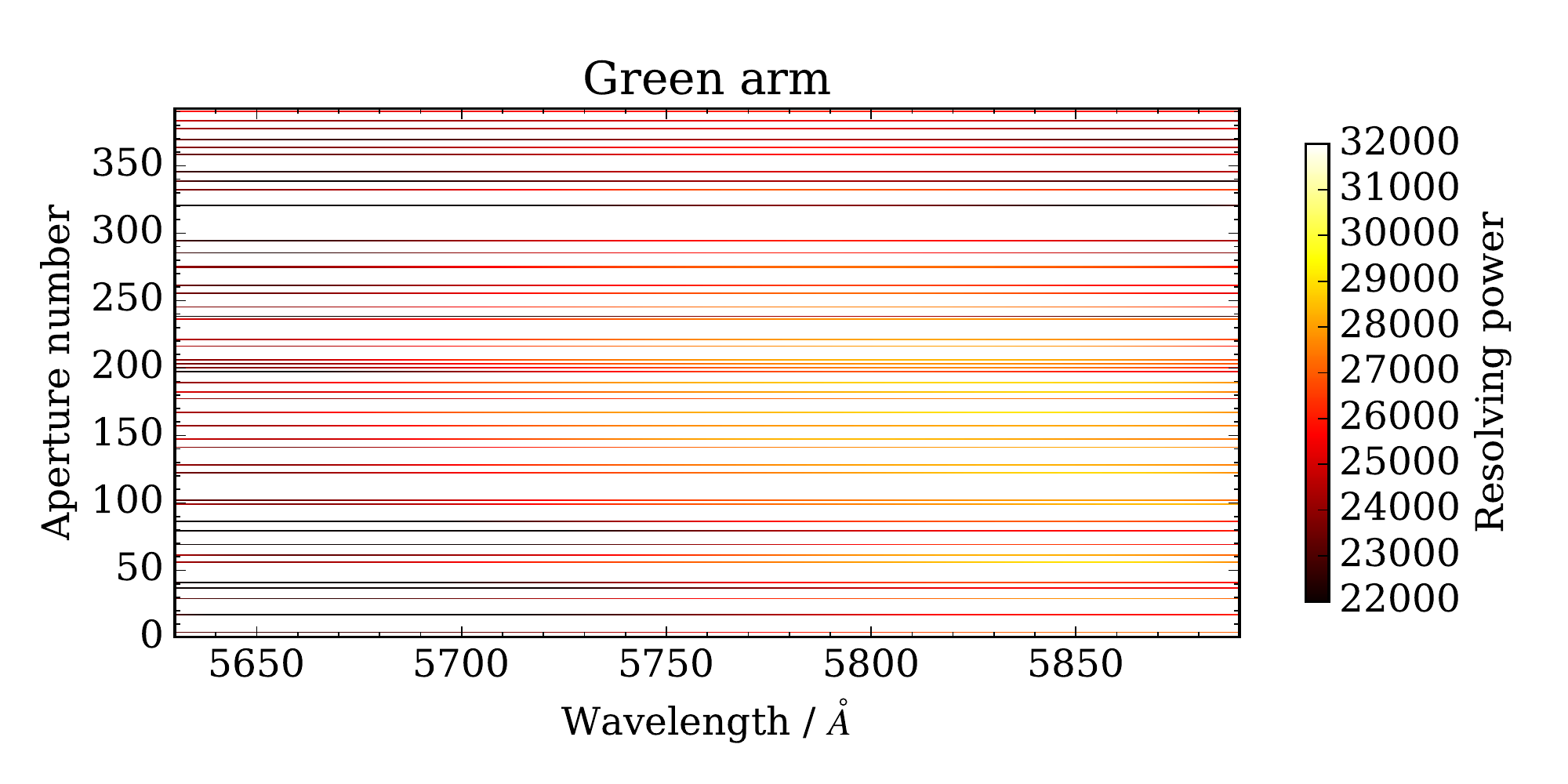}\\[-0.3cm]
\includegraphics[width=\columnwidth]{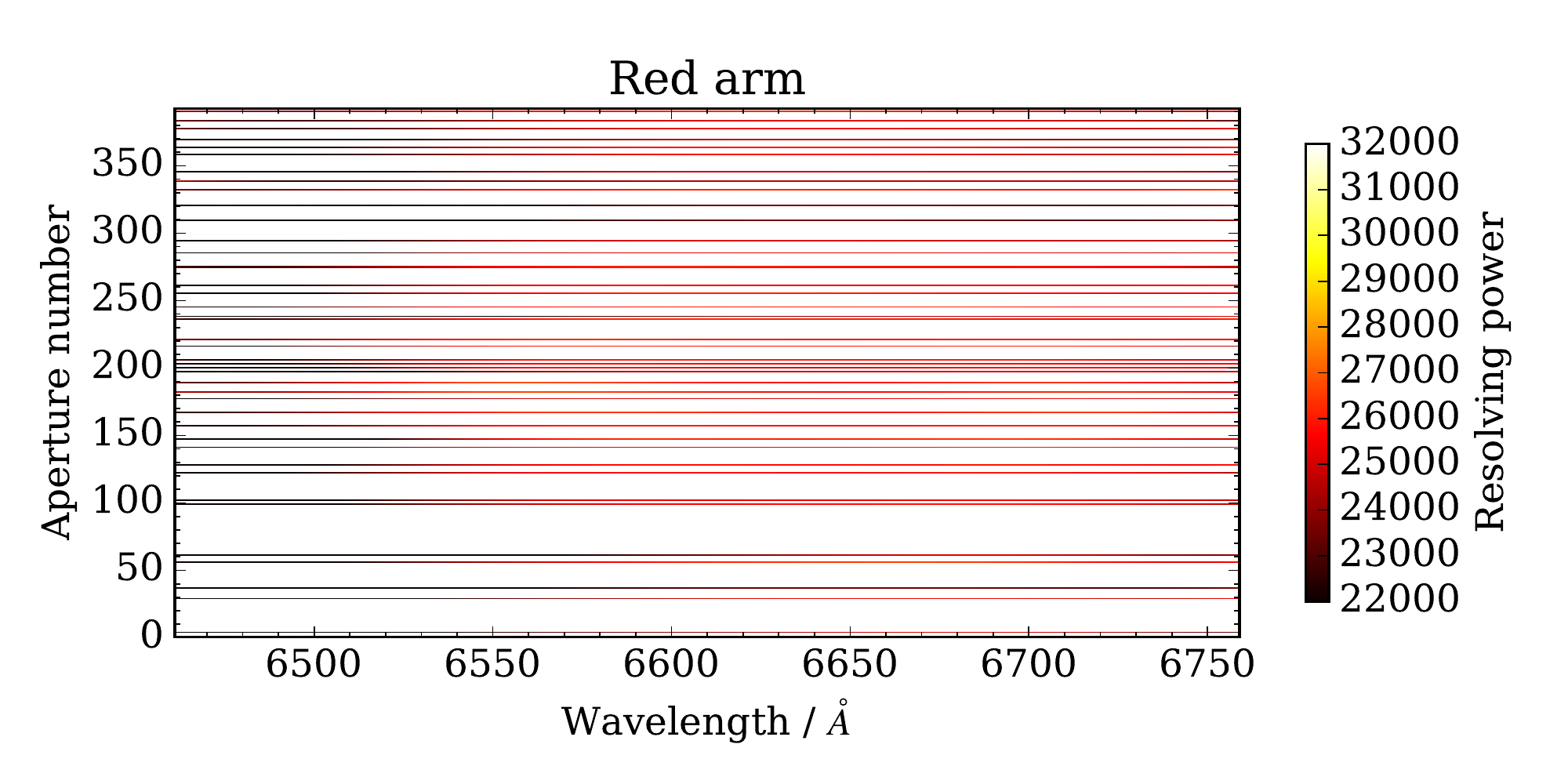}\\[-0.3cm]
\includegraphics[width=\columnwidth]{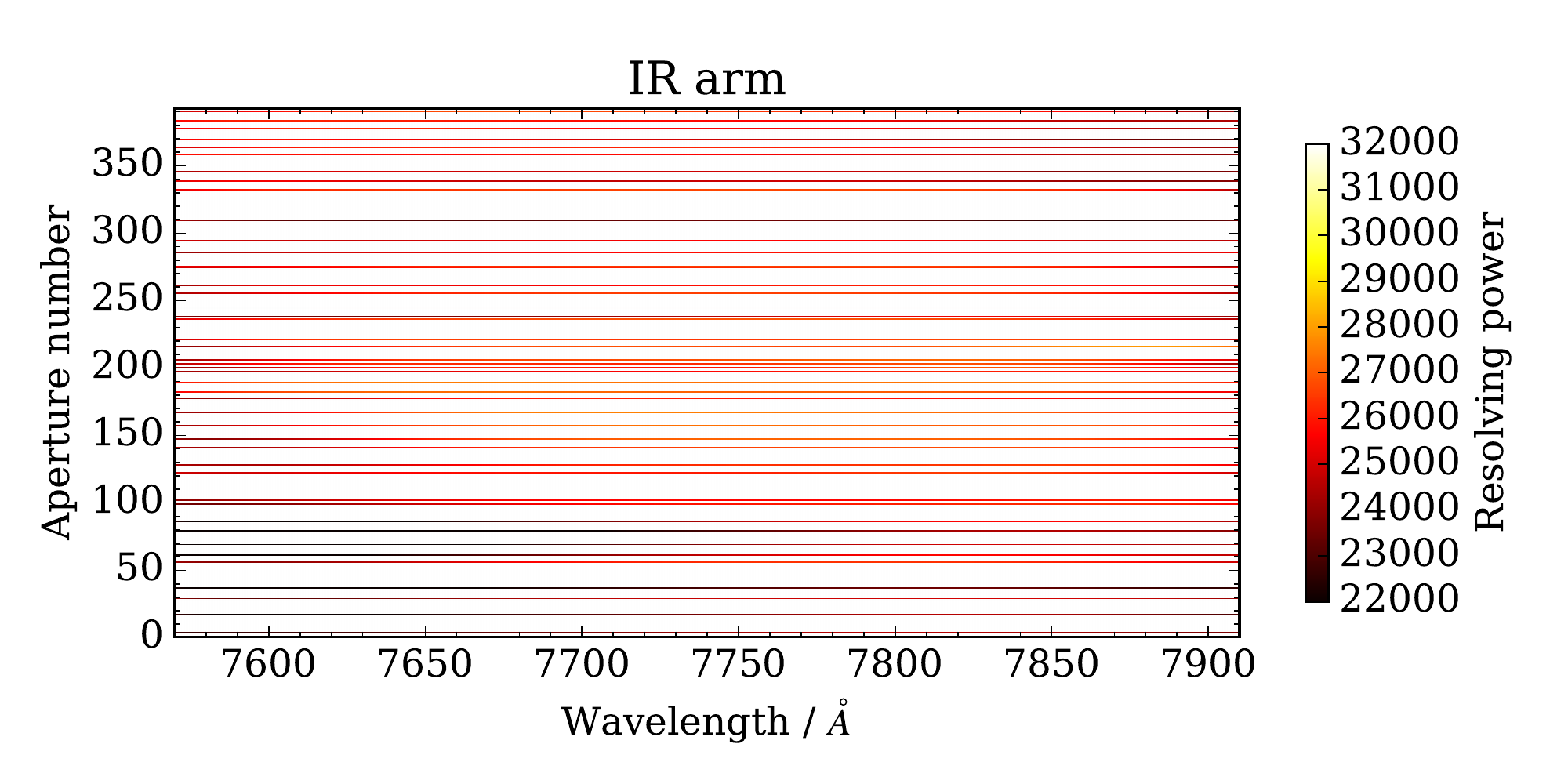}\\[-0.3cm]
\caption{Resolution maps for all three analysed arms with fibres from plate 0. Only 43 fibres were probed. White gaps represent fibres that were not probed.}
\label{fig:comb3}
\end{figure}

Figure \ref{fig:comb1} shows one such fit. Note that the central part of the peak is fitted equally well by a Gaussian and Voigt profile, while wings can only be fitted by a Voigt profile. This means that the wings give all the information about the Lorentzian part of the Voigt profile. The resolution in Figure \ref{fig:comb3} is measured as a FWHM of the Gaussian part of the Voigt profile. 

Figure \ref{fig:comb3} can be compared to Figures \ref{fig:res_map} and \ref{fig:res_map2}. While resolution measured via arc lines and the photonic comb is consistent in green and red arms, the shape is different in the IR arm. The wavy pattern along the wavelength axis that we see in Figure \ref{fig:res_map2} is absent in Figure \ref{fig:comb3}. We conclude that the part with lower resolution at around 7650~{\AA} that is making the pattern wavy is caused by the arc lines that are actually unresolved blends of several lines.

\section{Conclusions}

In this paper we presented the current state of the Iraf reduction pipeline for the GALAH survey. The code incorporates all major steps needed to extract a one dimensional spectrum from the acquired images and correct for known aberrations. Spectra are wavelength calibrated, but not flux calibrated, because this is not part of GALAH's science goals, and the chosen observing strategy therefore does not support it. The targeted precision of the calibrations and corrections that are made also driven by GALAH's scientific goals. The abundance pipelines require the wavelength calibration to be only slightly better than 1~$\mathrm{km\ s^-1}$, which is achieved, but the requirement is not surpassed. Therefore the GALAH project does not yet serve as a very precise radial velocity survey. The same principles guide the atmospheric parameters and radial velocity estimator -- it is designed to only give a first approximation of stellar atmospheric parameters that more advanced analysis pipelines can use.

Because the survey is still ongoing and observations will continue for some time, this will certainly not be the final version of this code. The observing strategy is expected to remain the same, but there is a possibility that new observing projects will be incorporated into the GALAH workflow. The GALAH survey is the first project to use the new HERMES spectrograph at the AAT. We therefore present some performance analysis (resolution maps) and give references (ThXe arc line lists) that may be useful to other users of HERMES. Apart from simulations and laboratory tests, results presented in this paper are the first use of the HERMES spectrograph in a real-life application.

Maintenance of the HERMES spectrograph is planned for the near future. This aims to fix problems with the slightly out-of-focus red arm CCD. The source of the vertical streaks will also be sought and possibly eliminated by replacing some optical elements \citep{sarah2016}. This will change the performance of the spectrograph, but should not impact the Iraf reduction pipeline too much. While it is expected that some fine tuning may be required after this maintenance, the overall process will remain the same. The resolution will also have to be measured again, as it might change after some optical elements are replaced. We plan to utilize an optical comb to do this job, because it proved to produce much more reliable and controlled spectra than the ThXe or ThAr lamp. In order to keep a consistent quality control, we plan to probe all fibres on both plates before and after the maintenance. We also plan to recalibrate the Xe line wavelengths with the optical comb to improve the wavelength calibration to a level where the precision of measured radial velocities will be limited by the spectral noise and not the wavelength calibration.

\section*{Acknowledgements}
We thank the anonymous referee for valuable comments and suggestions that improved the quality of the paper and its presentation.

The data in this paper were based on observations obtained at the Australian Astronomical Observatory as part of the GALAH survey (AAO Programs 2013B/13, 2014A/25, 2015A/19).

JK is funded by ARC grant DP150104667 awarded to JBH and TB.

This research has been supported in part by the Australian Research Council (ARC) funding schemes (grant numbers DP1095368, DP120101815, DP120101237, DP120104562, FS110200035 and FL110100012). SLM acknowledges the support of the ARC through grant DE140100598. DMN was supported by the ARC grant FL110100012.




\bibliographystyle{mnras}
\bibliography{bib}


\clearpage
\appendix

\section{{Th-Xe} linelists}
\label{sec:linelist2}

\begin{figure*}
\includegraphics[width=0.95\textwidth]{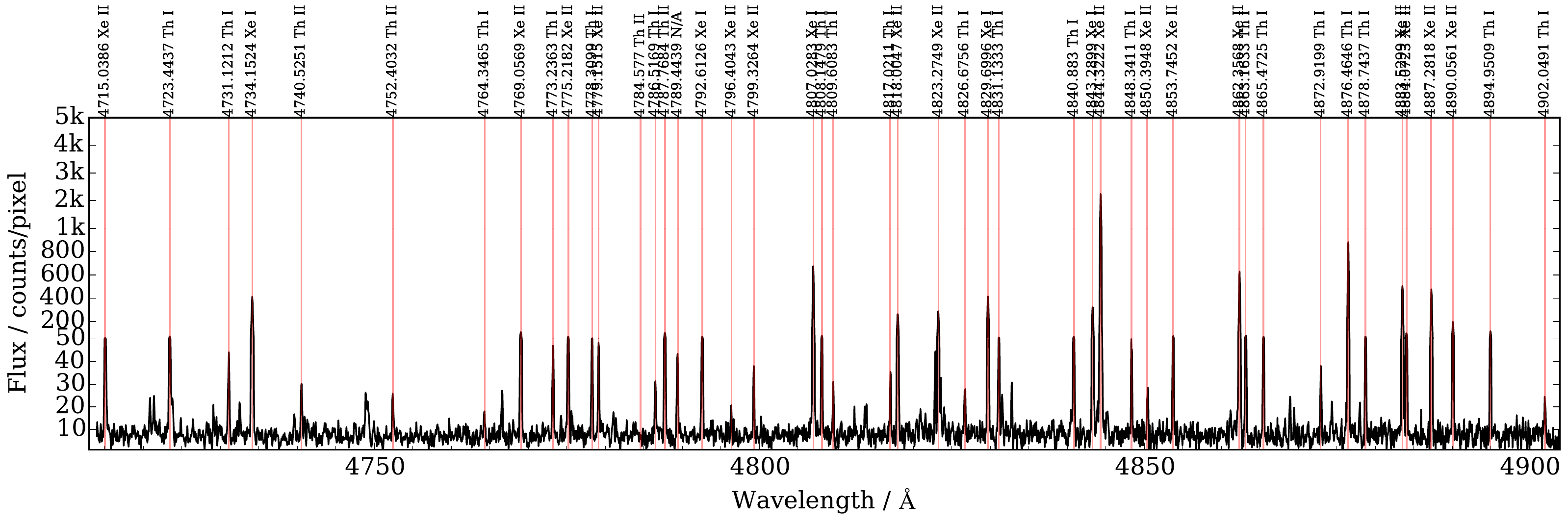}\\[-0.1cm]
\includegraphics[width=0.95\textwidth]{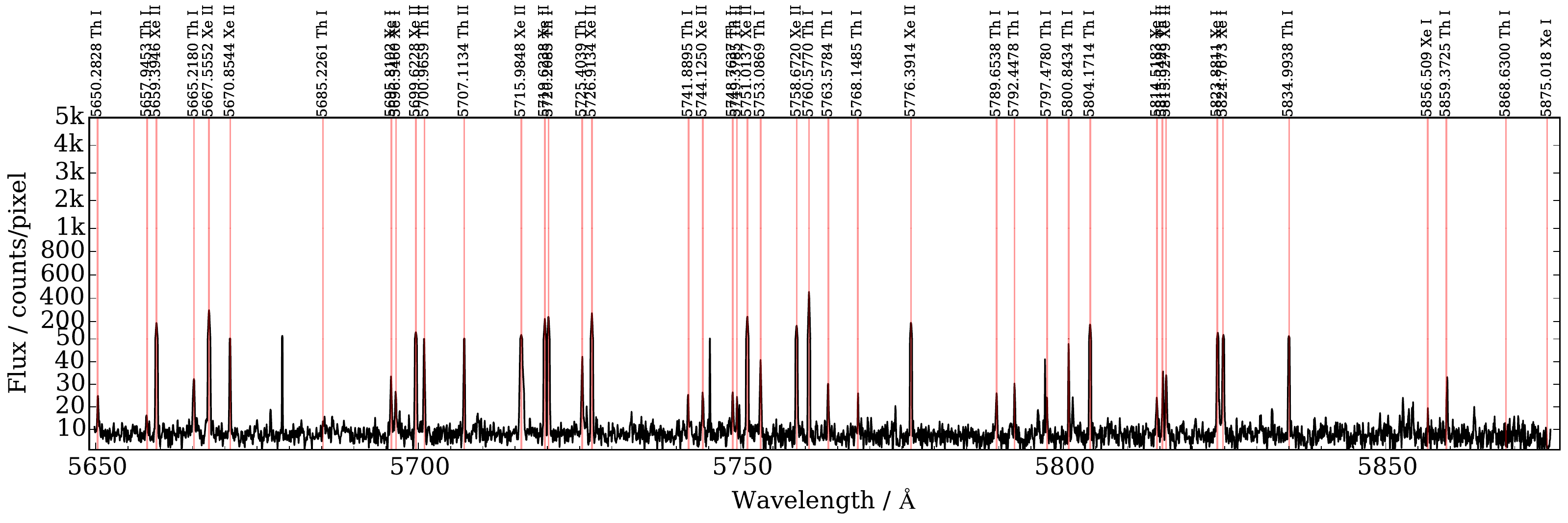}\\[-0.1cm]
\includegraphics[width=0.95\textwidth]{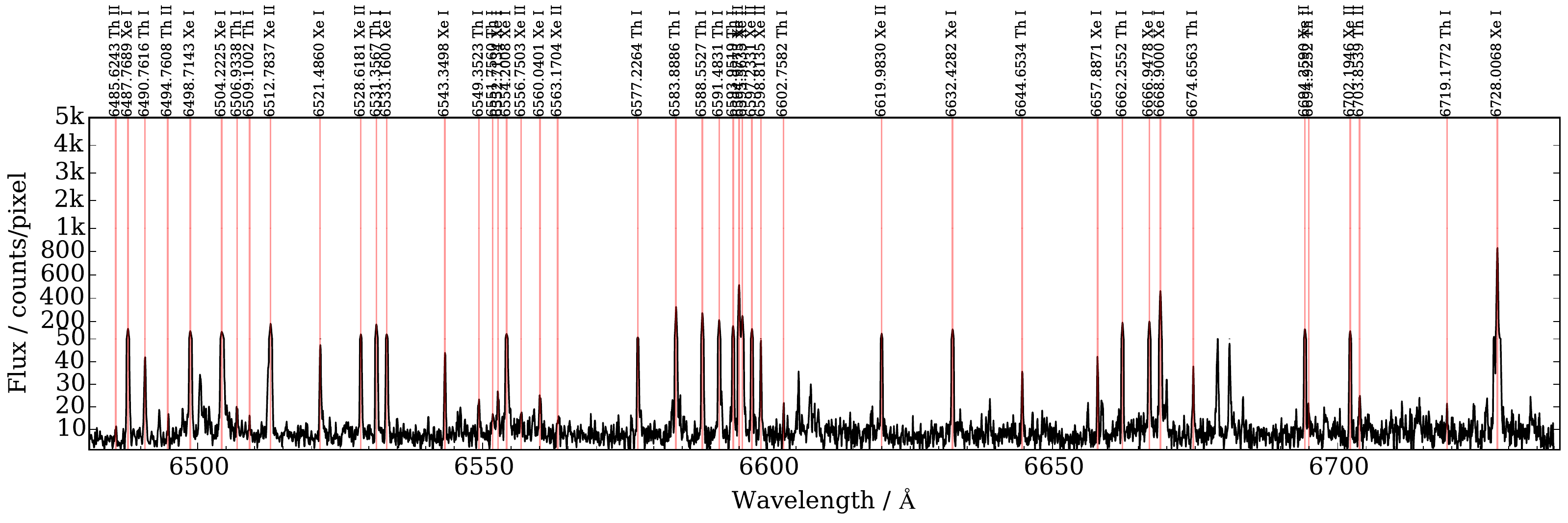}\\[-0.1cm]
\includegraphics[width=0.95\textwidth]{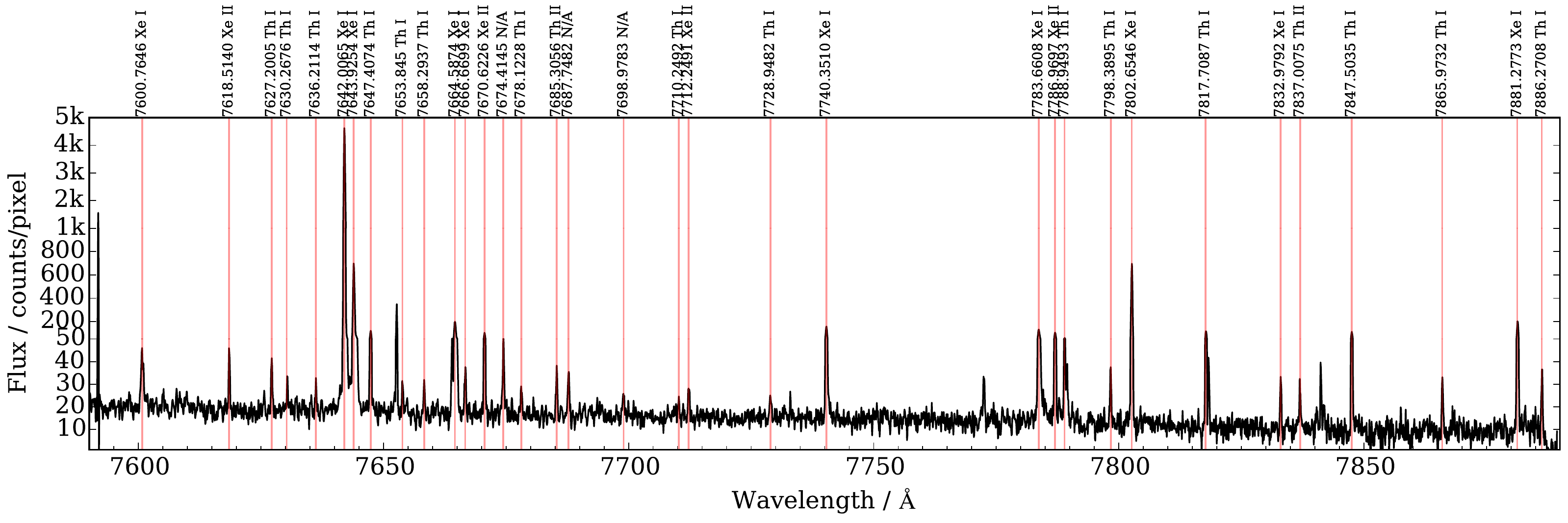}\\[-0.1cm]
\caption{Typical arc spectra with marked lines included in the line lists. Blue, green, red and IR arms follow top to bottom. Some strong, but unmarked lines are cosmetic errors and some unmarked lines have peculiar shapes, so the did not make it into the line lists. The wavelengths are taken from our line list. There is a lot of dynamic range in the flux, so we used a non-linear flux scale in all four panels.}
\label{fig:arc_spectra}
\end{figure*}

The tables below give the wavelengths of Thorium and Xe lines used in our line lists. Although 5 decimal places are given for each wavelength, we do not suggest such accuracy. See section \ref{sec:wavs} for the estimation of the precision of the method we used to get these wavelengths. All the lines but one are matched to the lines in the literature, although the match is sometimes ambiguous. Figure \ref{fig:arc_spectra} shows a typical arc spectrum with matched arc lines for all four arms.

\begin{table}
\begin{tabular}{lllll}
\multicolumn{5}{c}{Blue arm}\\\hline\hline
Our $\lambda$ & Literature $\lambda$ & Line & Source & Note\\\hline
4715.03863 & 4715.18 & Xe II & a &$\dagger$\\
4723.44378 & 4723.4382 & Th I & b& \\
4731.12127 & 4731.1955 & Th I & b & \\
4734.15242 & 4734.152 & Xe I & c & \\
4740.52510 & 4740.5292 & Th II & b& \\
4752.40323 & 4752.4141 & Th II & b& \\
4764.34652 & 4764.3463 & Th I & b& \\
4769.05694 & 4769.05 & Xe II & a & \\
4773.23634 & 4773.241 & Th I & b& \\
4775.21822 & 4775.18  & Xe II & a & \\
4778.30990 & 4778.294 & Th I & b& \\
4779.15157 & 4779.18  & Xe II & a & \\
4784.57700 & 4784.5667 & Th II & d & \\
4786.51691 & 4786.531 & Th I & b& \\
4787.76840 & 4787.7462 & Th II & d &$\clubsuit$\\
4789.44399 & & Th I, Th II & e, b & $\clubsuit\clubsuit$\\
4792.61260 & 4792.619 & Xe I & c & \\
4796.40438 & 4796.48  & Xe II & a & \\
4799.32645 & 4799.45  & Xe II & a &$\ast$\\
4807.02831 & 4807.02  & Xe I & c & \\
4808.14795 & 4808.1337 & Th I & b& \\
4809.60832 & 4809.614 & Th I & b& \\
4817.02113 & 4817.0206 & Th I & b& \\
4818.00478 & 4818.02  & Xe II & a & \\
4823.27491 & 4823.35  & Xe II & a &$\ast\ast$\\
4826.67568 & 4826.7004 & Th I & b& \\
4829.69964 & 4829.71  & Xe I & c & \\
4831.13337 & 4831.1213 & Th I & b& \\
4840.88300 & 4840.8492 & Th I & b& \\
4843.28996 & 4843.29  & Xe I & c & \\
4844.32227 & 4844.33  & Xe II & a & \\
4848.34116 & 4848.3625 & Th I & b& \\
4850.39484 & 4850.23  & Xe II & a &$\ast$\\
4853.74521 & 4853.77 & Xe II & a & \\
4862.35682 & 4862.45  & Xe II & a &$\ast$\\
4863.16331 & 4863.1724 & Th I & b& \\
4865.47257 & 4865.4775 & Th I & b & \\
4872.91990 & 4872.9169 & Th I & b& \\
4876.46465 & 4876.4949 & Th I & b& \\
4878.74379 & 4878.733 & Th I & b& \\
4883.52998 & 4883.53  & Xe II & a & \\
4884.07236 & 4884.07  & Xe II & a &$\ast\ast$\\
4887.28188 & 4887.29  & Xe II & a & \\
4890.05617 & 4890.09  & Xe II & a & \\
4894.95091 & 4894.9551 & Th I & b& \\
4902.04918 & 4902.0545 & Th I & b & \\\hline
\end{tabular}
\caption{Lines from our blue arm linelist and wavelengths from the literature. First column gives the wavelengths we assumed for our linelist. Second column gives wavelengths of the same lines as found in the literature. Third column shows the element and its ionization state responsible for the line. Literature source and notes are in the remaining two columns. Literature sources: a: \citet{hansen1987}, b: \citet{palmer1983}, c: \citet{meggers1933}, d: \citet{lovis2007}, e: \citet{redman2014}, f: \citet{humphreys1939}, g: \citet{meggers1934}, h: \citet{zalubas1976}, j: \citet{ahmed1998}, k: \citet{zalubas1974}. Notes: $\dagger$: Wavelengths don't match, but there is a line nearby in the literature. Hence we put the wavelength of the nearest line into the table. $\ast$: Line is marked hazy in the literature. $\ast\ast$: line is marked very hazy in the literature. $\ddagger$: The wavelength does not match, but there are two lines nearby, one of Xe II and one of Th I. $\clubsuit$: This is a blend of two equally strong lines. The average wavelength is given in the table. $\clubsuit\clubsuit$: This is probably a blend of two Thorium lines. They have different strength in the literature and come from different ionization stages. It is therefore impossible to  calculate a precise average wavelength.}
\label{tab:linelist}
\end{table}

\begin{table}
\begin{tabular}{lllll}
\multicolumn{5}{c}{Green arm}\\\hline\hline
Our $\lambda$ & Literature $\lambda$ & Line & Source & Note\\\hline
5650.28289 & 5650.3276 & Th I & b & \\
5657.94536 & 5657.9255 & Th I & b& \\
5659.39468 & 5659.38  & Xe II & f & \\
5665.21802 & 5665.1799 & Th I & b& \\
5667.55524 & 5667.56  & Xe II & f & \\
5670.85448 & 5670.91  & Xe II & f & \\
5685.22615 & 5685.1921 & Th I & b& \\
5695.81028 & 5695.75  & Xe I & c & \\
5696.54602 & 5696.477 & Xe I & g & \\
5699.62286 & 5699.61  & Xe II & f & \\
5700.96596 & 5700.9176 & Th II & b& \\
5707.11348 & 5707.1033 & Th II & b& \\
5715.98484 & 5716.10  & Xe II & f & \\
5719.62383 & 5719.61  & Xe II & a & \\
5720.20852 & 5720.1828 & Th I & b& \\
5725.40398 & 5725.3885 & Th I & b& \\
5726.91340 & 5726.91  & Xe II & f & \\
5741.88955 & 5741.82881 & Th I & e & \\
5744.12509 & 5744.2   & Xe II & a & \\
5748.76371 & 5748.7412 & Th I & b& \\
5749.37855 & 5749.3883 & Th II & b& \\
5751.01375 & 5751.03  & Xe II & f & \\
5753.08690 & 5753.0265 & Th I & b& \\
5758.67208 & 5758.65  & Xe II & f & \\
5760.57706 & 5760.5508 & Th I & b& \\
5763.57841 & 5763.529 & Th I & b& \\
5768.14854 & 5768.1812 & Th I & b& \\
5776.39142 & 5776.39  & Xe II & f & \\
5789.65389 & 5789.6451 & Th I & b& \\
5792.44789 & 5792.4304 & Th I & b& \\
5797.47803 & 5797.3194 & Th I &  b&$\dagger$\\
5800.84347 & 5800.8297 & Th I & b& \\
5804.17143 & 5804.1412 & Th I & b& \\
5814.51830 & 5814.505 & Xe I & g & \\
5815.34288 & 5815.4219 & Th II & b &$\dagger$\\
5815.92798 & 5815.96  & Xe II & f & \\
5823.88112 & 5823.89  & Xe I & c & \\
5824.76730 & 5824.80  & Xe I & c & \\
5834.99388 & 5835.167 & Th I & d &$\dagger$\\
5856.50900 & 5856.509 & Xe I & g & \\
5859.37251 & 5859.348 & Th I & h & \\
5868.63003 & 5868.37475 & Th I & e &$\dagger$\\
5875.01800 & 5875.02  & Xe I & c & \\\hline
\end{tabular}
\addtocounter{table}{-1}
\caption{Table continues for the green arm.}
\end{table}

\begin{table}
\begin{tabular}{lllll}
\multicolumn{5}{c}{Red arm}\\\hline\hline
Our $\lambda$ & Literature $\lambda$ & Line & Source & Note\\\hline
6485.62436 & 6485.3759 & Th II & b &$\dagger$\\
6487.76893 & 6487.76  & Xe I & c & \\
6490.76162 & 6490.7372 & Th I &b & \\
6494.76087 & 6494.981 & Th II & d &$\dagger$\\
6498.71437 & 6498.72  & Xe I & c & \\
6504.22250 & 6504.18  & Xe I & c & \\
6506.93387 & 6506.9863 & Th I & b& \\
6509.10027 & 6509.0503 & Th I & b& \\
6512.78374 & 6512.83  & Xe II & f & \\
6521.48606 & 6521.508 & Xe I & g & \\
6528.61811 & 6528.65  & Xe II & f & \\
6531.35671 & 6531.3418 & Th I & b& \\
6533.16006 & 6533.16  & Xe I & c & \\
6543.34982 & 6543.360 & Xe I & g & \\
6549.35230 & 6549.820 & Th I & h &$\dagger$\\
6551.75606 & 6551.7055 & Th I & b& \\
6552.71340 & 6552.79  & Xe I & j &$\clubsuit$\\
6554.20081 & 6554.196 & Xe I & g & \\
6556.75037 & 6556.70  & Xe II & a & \\
6560.04011 & 6559.97  & Xe I & c & \\
6563.17049 & 6563.19  & Xe II & a & \\
6577.22645 & 6577.2146 & Th I & b& \\
6583.88861 & 6583.906 & Th I & b& \\
6588.55275 & 6588.5396 & Th I &b & \\
6591.48314 & 6591.4845 & Th I & b& \\
6593.95197 & 6593.9391 & Th I & b& \\
6594.97730 & 6595.01  & Xe II & f & \\
6595.56355 & 6595.56  & Xe I & c & \\
6597.23314 & 6597.25  & Xe II & f & \\
6598.81354 & 6598.84  & Xe II & f & \\
6602.75824 & 6602.7620 & Th I & b & \\
6619.98306 & 6620.02  & Xe II & a & \\
6632.42826 & 6632.464 & Xe I & g & \\
6644.65341 & 6644.6700 & Th I & b & \\
6657.88712 & 6657.92  & Xe I & c & \\
6662.25522 & 6662.2686 & Th I & b& \\
6666.94783 & 6666.965 & Xe I & g & \\
6668.90004 & 6668.92  & Xe I & c & \\
6674.65633 & 6674.6969 & Th I & b& \\
6694.25904 & 6694.32  & Xe II & f & \\
6694.92523 & 6695.181 & Th I & d &$\dagger$\\
6702.19461 & 6702.25  & Xe II & a & \\
6703.85395 & 6703.8966 & Th II & b & \\
6719.17725 & 6719.19943 & Th I & e & \\
6728.00680 & 6728.01  & Xe I & c & \\\hline
\end{tabular}
\addtocounter{table}{-1}
\caption{Table continues for the red arm.}
\end{table}

\begin{table}
\begin{tabular}{lllll}
\multicolumn{5}{c}{IR arm}\\\hline\hline
Our $\lambda$ & Literature $\lambda$ & Line & Source & Note\\\hline
7600.76466 & 7600.77  & Xe I & c & \\
7618.51401 & 7618.57  & Xe II & f & \\
7627.20055 & 7627.1749 & Th I & b& \\
7630.26763 & 7630.3106 & Th I & b& \\
7636.21145 & 7636.17473 & Th I & e & \\
7642.00652 & 7642.02  & Xe I & c & \\
7643.92543 & 7643.91  & Xe I & c & \\
7647.40741 & 7647.3794 & Th I & b& \\
7653.84500 & 7653.8284 & Th I & b& \\
7658.29378 & 7658.3202 & Th I & b& \\
7664.58740 & 7664.56  & Xe I & c & \\
7666.66991 & 7666.61  & Xe I & c & \\
7670.62263 & 7670.66  & Xe II & f & \\
7674.41459 & N/A & N/A & N/A & \\
7678.12287 & 7678.1267 & Th I & b& \\
7685.30565 & 7685.3075 & Th II & b& \\
7687.74823 & N/A & N/A & N/A & \\
7698.97833 & N/A & N/A & N/A & \\
7710.24928 & 7710.2689 & Th I & b &\\
7712.24919 & 7712.42,7712.405 & Th I, Xe II & a &$\ddagger$\\
7728.94825 & 7728.9509 & Th I & b &\\
7740.35102 & 7740.31  & Xe I & c & \\
7783.66085 & 7783.66  & Xe I & c & \\
7786.96975 & 7787.04  & Xe II & f & \\
7788.94932 & 7788.9342 & Th I & b& \\
7798.38957 & 7798.35789 & Th I & e & \\
7802.65468 & 7802.65  & Xe I & c & \\
7817.70870 & 7817.7669 & Th I & b& \\
7832.97925 & 7832.98 & Xe I & c& \\
7837.00751 & 7836.703 & Th II & k &$\dagger$\\
7847.50351 & 7847.5394 & Th I & b& \\
7865.97327 & 7865.9698 & Th I & b& \\
7881.27733 & 7881.32  & Xe I & c & \\
7886.27084 & 7886.283 & Th I & b& \\\hline
\end{tabular}
\addtocounter{table}{-1}
\caption{Table continues for the IR arm.}
\end{table}

\bsp
\label{lastpage}
\end{document}